\documentclass[useAMS,usenatbib]{mn2e}
\bibpunct{[}{]}{,}{n}{}{;}

\usepackage{amssymb,amsmath,upgreek}

\usepackage{epsfig,graphicx,times,subfigure}
\newcommand{\kms}{{km~s$^{-1}$}}

\newcommand{\msun}{M$_{\odot}$}

\newcommand{\lsun}{L$_{\odot}$}

\newcommand{\aaps}{A\&A}

\newcommand{\apj}{ApJ}
\newcommand{\mnras}{MNRAS}
\newcommand{\araa}{ARA\&A}
\newcommand{\apjs}{ApJSup.Series}

\def\deg{\hbox{$^\circ$}}

\title[Supernova 2009md]
{SN 2009md: Another faint supernova from a low mass progenitor}

\author[M. Fraser et al.]
{M. Fraser$^{1}$\thanks{E-mail:mfraser02@qub.ac.uk}, M. Ergon$^{2}$, J.J. Eldridge$^{3}$, S. Valenti$^{1}$, A. Pastorello${^1}$, J. Sollerman$^{2}$, 
\newauthor  S.J. Smartt$^{1}$, I. Agnoletto$^{5}$, I. Arcavi$^{6}$, S. Benetti$^{5}$, M.-T. Botticella$^{1}$, F. Bufano$^{5}$,
\newauthor A. Campillay$^{7}$, R.M. Crockett$^{8}$, A. Gal-Yam$^{6}$,  E. Kankare$^{9,10}$, G. Leloudas$^{12}$, K. Maguire$^{1,8}$,
\newauthor  S. Mattila$^{10,11}$, J.R. Maund$^{12}$, F. Salgado$^{7}$, A. Stephens$^{13}$, S. Taubenberger$^{14}$, M. Turatto$^{5}$ 
\\$^{1}$Astrophysics Research Center, School of Mathematics and Physics, Queen's University Belfast, Belfast, BT7 1NN, Northern Ireland
\\$^{2}$Oskar Klein Centre, Department of Astronomy, AlbaNova, Stockholm University, 106 91 Stockholm, Sweden
\\$^{3}$Institute of Astronomy, University of Cambridge, Madingley Road, Cambridge, CB3 0HA, UK
\\$^{4}$Dipartimento di Astronomia, Universit\'a di Padova, Vicolo dell'Osservatorio 3, 35122 Padova, Italy
\\$^{5}$INAF-Osservatorio Astronomico di Padova, Vicolo dell'Osservatorio 5, 35122 Padova, Italy
\\$^{6}$Weizmann Institute of Science, Rehovot 76100, Israel
\\$^{7}$Las Campanas Observatory, Carnegie Observatories, Casilla 601, La Serena, Chile
\\$^{8}$Department of Physics (Astrophysics), University of Oxford, Keble Road, Oxford OX1 3RH, UK
\\$^{9}$Nordic Optical Telescope, Apartado 474, E-38700 Santa Cruz de La Palma, Spain 
\\$^{10}$Tuorla Observatory, Department of Physics and Astronomy, University of Turku, V\"ais\"al\"antie 20, FI-21500, Finland 
\\$^{11}$Stockholm Observatory, Department of Astronomy, AlbaNova University Center, SE-106 91 Stockholm, Sweden 
\\$^{12}$Dark Cosmology Centre, Juliane Maries Vej 30, 2100 Copenhagen \O\ Denmark 
\\$^{13}$Gemini Observatory, 670 North AÕohoku Place, Hilo, HI 96720, USA
\\$^{14}$Max-Planck-Institut fŸr Astrophysik, Karl-Schwarzschild-Str. 1, D-85748 Garching, Germany
}

\begin{document}

\date{Submitted to Monthly Notices of the Royal Astronomical Society}

\pagerange{\pageref{firstpage}--\pageref{lastpage}} \pubyear{}

\maketitle

\label{firstpage}

\begin{abstract}
We present adaptive optics imaging of the core collapse supernova (SN) 2009md, which we use together with archival \emph{Hubble Space Telescope} data to identify a coincident progenitor candidate. We find the progenitor to have an absolute magnitude of $V = -4.63^{+0.3}_{-0.4}$ mag and a colour of $V-I = 2.29^{+0.25}_{-0.39}$ mag, corresponding to a progenitor luminosity of log $L$/\lsun $\sim4.54\pm0.19$ dex. Using the stellar evolution code STARS, we find this to be consistent with a red supergiant progenitor with $M  = 8.5_{-1.5}^{+6.5}$ \msun. The photometric and spectroscopic evolution of SN 2009md is similar to that of the class of sub-luminous Type IIP SNe; in this paper we compare the evolution of SN 2009md primarily to that of the sub-luminous SN 2005cs. We estimate the mass of $^{56}$Ni ejected in the explosion to be $(5.4\pm1.3) \times 10^{-3}$ \msun\ from the luminosity on the radioactive tail, which is in agreement with the low $^{56}$Ni masses estimated for other sub-luminous Type IIP SNe. From the lightcurve and spectra, we show the SN explosion had a lower energy and ejecta mass than the normal Type IIP SN 1999em. We discuss problems with stellar evolutionary models, and the discrepancy between low observed progenitor luminosities (log $L$/\lsun $\sim4.3-5$ dex) and model luminosities after the second-dredge-up for stars in this mass range, and consider an enhanced carbon burning rate as a possible solution. In conclusion, SN 2009md is a faint SN arising from the collapse of a progenitor close to the lower mass limit for core-collapse. This is now the third discovery of a low mass progenitor star producing a low energy explosion and low $^{56}$Ni ejected mass, which indicates that such events arise from the lowest end of the mass range that produces a core-collapse supernova ($7-8$ \msun).
\end{abstract}


\section{Introduction}
\label{s1}

Core-collapse SNe mark the endpoint of stellar evolution for stars more massive than $\sim$8 \msun\ (Poelarends et al. \citealp{Poe08}; Siess \citealp{Sie07}; Heger et al. \citealp{Heg03}; Eldridge \& Tout \citealp{Eld04}). When a star more massive than this exhausts its nuclear fuel, it can no longer support itself against gravitational collapse. The ensuing collapse, and subsequent explosion of the progenitor star gives rise to the supernova's spectacular display. The fact that the progenitors of core-collapse SNe are massive, and hence luminous, gives a realistic prospect of successfully recovering them in high resolution archival data (see Smartt \citealp{Sma09a} for a review). Since the first identification of a core collapse SN progenitor, for SN 1987A in the Large Magellanic Cloud (West, Reipurth \& Jorgenson \citealp{Wes87}; White \& Malin \citealp{Whi87}), there have been numerous direct detections of SN progenitors (for example, SN 1993J, Aldering, Humphreys \& Richmond \citealp{Ald94}; SN 2004A, Hendry et al. \citealp{Hen06}; SN 2004et, Li et al. \citealp{Li05}; Crockett et al. \citealp{Cro10}; SN 2005cs, Maund, Smartt \& Danziger \citealp{Mau05}; Li et al. \citealp{Li06}; SN 2008cn, Elias-Rosa et al. \citealp{Eli09}). Of particular interest is the family of sub-luminous Type IIP SNe (Pastorello et al. \citealp{Pas04}), which are typified by fainter absolute magnitudes, lower expansion velocities and smaller ejected $^{56}$Ni masses than the canonical Type IIP SNe. In recent years, there has been some debate in the literature as to the precise nature of these events, and whether they represent the collapse of a low mass ($\sim$9 \msun) star with a O-Ne-Mg core (Kitaura, Janka \& Hillebrandt \citealp{Kit06}), or a high mass ($\sim$25 \msun) star with the formation of a black hole, possibly by the fallback of material onto an accreting proto-neutron star (Turatto et al. \citealp{Tur98}; Zampieri et al. \citealp{Zam03}). The recovery of progenitors in archival data can help determine which scenario is (or whether both, or neither, are) correct.

SN 1997D was the prototype of this class of sub-luminous Type IIP SNe; at discovery the SN had line velocities of $\sim $1200 \kms, at later epochs a low ejected nickel mass was measured from the radioactive tail (Benetti et al. \citealp{Ben01}). Turatto et al. (\citealp{Tur98}) claimed SN 1997D resulted from the collapse of a high mass (26 \msun) progenitor. This was based however on SN modeling, which gives consistently higher masses than found from progenitor modeling (Maguire et al. \citealp{Mag10a}). Since the discovery of SN 1997D about a dozen sub-luminous Type IIP SNe have been discovered and classified (Pastorello et al. \citealp{Pas04}; Spiro \& Pastorello \citealp{Spi09}). Out of these, four have either progenitor detections or useful upper mass limits. SN 2005cs was the first sub-luminous Type IIP SN for which a $6-8$ \msun\ progenitor was detected (Maund, Smartt \& Danziger \citealp{Mau05}; Li et al. \citealp{Li06}; Eldridge, Mattila \& Smartt \citealp{Eld07}). The SN displayed a low expansion velocity, and had a faint absolute magnitude of $V = -14.75$ mag (Pastorello et al. \citealp{Pas06}, \citealp{Pas09}) and a low $^{56}$Ni mass. For SN 2008bk (Pignata et al. in prep.) another low mass, red supergiant progenitor of $8.5\pm1.0$ \msun was identified by Mattila et al. (\citealp{Mat08}) in optical and near-infrared pre-explosion imaging. The sub-luminous SNe for which we have a progenitor mass limit are SN 1999br (Filippenko \citealp{Fil99}; Pastorello et al. \citealp{Pas04}), where Maund \& Smartt (\citealp{Mau05}) give a progenitor mass $<$ 12 \msun\ (more recently revised to $M< 15$ \msun\ by Smartt et al. \citealp{Sma09b}), and SN 2006ov, where a progenitor limit $M <10$ \msun\ was set by Crockett et al. (\citealp{Cro10}). From the low luminosities of the progenitors, we can rule out very massive stars, and bright, low mass, super-asymptotic giant branch (SAGB) stars as the source of a significant fraction of these events.

In light of this, the detection of the progenitor of another sub-luminous Type IIP, coupled with the detailed monitoring and follow-up observations needed to understand the explosion, is of considerable interest. Understanding the nature of faint SNe, and how stars at the lower extremum of Type II SN progenitor masses give rise to them can even help shed light on other questions, such as the upper mass limit for the formation of SAGB stars (eg. Pumo et al. \citealp{Pum09}; Siess \citealp{Sie07}). Sub-luminous SNe may also nucleosynthesize different relative fractions of elements, and so influence the observed abundances in subsequent generations of stars (Tsujimoto \& Shigeyama \citealp{Tsu03}).

\begin{figure}
\includegraphics[width=0.475\textwidth,angle=0]{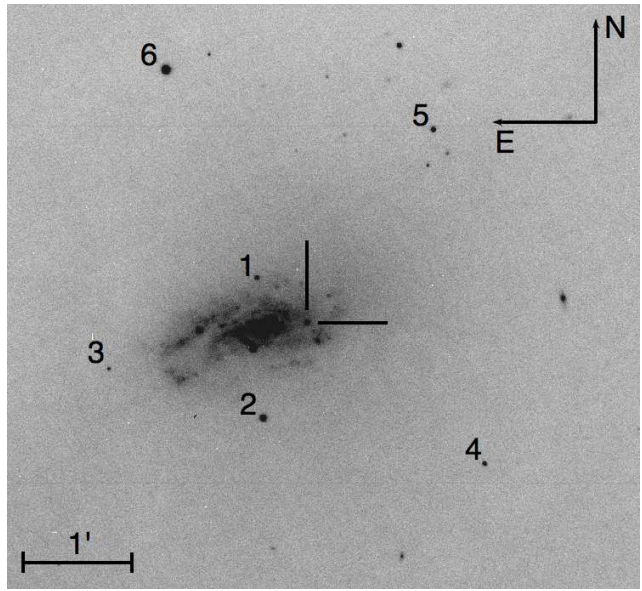}
\caption{5s unfiltered acquisition image of NGC 3389, obtained on 2009 December 07 with the Nordic Optical Telescope+ALFOSC. The location of SN 2009md is indicated with cross marks, sequence stars used to calibrate optical photometry are numbered according to Table \ref{t_seq}.}
\label{f1}
\end{figure}

SN 2009md was found at an unfiltered magnitude of 16.5 mag by K. Itagaki (Nakano \citealp{Nak09}) in the spiral galaxy NGC 3389 on 4 December 2009. The location of the SN is shown in Fig. \ref{f1}.  Sollerman et al. (\citealp{Sol09}) obtained a spectrum of the SN on 7 December 2009 with the Nordic Optical Telescope (NOT) + ALFOSC, and classified it as a young Type IIP. To constrain the explosion epoch, we used a 120 s $i'$ band image of NGC 3389 from P.-A. Duc, which was taken on 19 November 2010 with the Canada-France-Hawaii Telescope (CFHT) + MegaCam. The data were reduced by the {\sc elixir} pipeline; the SN is not visible in the image at this epoch to a limiting magnitude of $i' > 23.4$.

We also used the SN spectrum comparison tool {\sc gelato} (Harutyunyan et al. \citealp{Har08}) to compare the earliest spectrum available of SN 2009md, from 7 December 2009, to those of other Type II SNe, finding matches with Type II SNe at phases of +3 to +10 days. We have hence adopted an explosion epoch of 2009 November 27 (MJD 55162), with an uncertainty of $\pm 8$ days. This epoch will be used as a reference throughout the paper.

The rest of this paper is laid out as follows: In Section \ref{s2} we discuss the archival data covering the location of SN 2009md, our progenitor detection and analysis. We also characterize the host galaxy, and estimate metallicity and distance. Section \ref{s3} deals with photometry and spectra of the SN itself, we also present a bolometric lightcurve for SN 2009md and an estimate of the ejected $^{56}$Ni mass. In Section \ref{s4} we present further discussion and analysis of the SN. We verify the distance to NGC 3389 with the standard candle method for Type IIP SNe, discuss how changes to stellar evolutionary models may help ameliorate the discrepancy between the observed low luminosity of the progenitor of SN 2009md and the final luminosities found from models, estimate the explosion energy, and discuss the properties of the family of sub-luminous Type IIP SNe.

\section{Host galaxy and progenitor candidate}
\label{s2}
\subsection{NGC 3389}
\label{s2f}

The host galaxy, NGC 3389, has a distance modulus $\mu = 31.85 \pm 0.15$ mag from its recessional velocity (1712 km s$^{-1}$, value from NED\footnote{$H_0 = 73$ \kms Mpc$^{-1}$}) after correcting for infall on Virgo, GA and Shapley. Terry, Paturel \& Ekholm (\citealp{Ter02}) find $\mu = 31.43 \pm 0.14$ mag from the ``sosies'' method, while the Tully-Fisher relation gives $\mu = 31.76 \pm 0.8$ mag (Tully \citealp{Tul88}). We took the mean of the three measurements (weighted by the inverse of the associated uncertainty for each method) to obtain a distance modulus $\mu = 31.64 \pm 0.21$ mag. The error in the distance modulus was calculated as the range of the three values about the mean.

Ideally, to estimate the metallicity at the SN location we would use a measurement of the ratio of the [O{\sc ii}] and [O{\sc iii}] to H$\beta$ lines in a H{\sc ii} region close to the SN (Bresolin \citealp{Bre08}). Unfortunately, we do not have such a measurement for NGC 3389.\footnote{We note that there is published photometry of H{\sc ii} regions in NGC 3389 (Abdel-Hamid, Lee \& Notni \citealp{Abd03}), with two regions close to the position of SN 2009md. The closer region, H14, has a projected distance $\sim$2\arcsec, the more distant, H16, has a distance of $\sim$ 3.5\arcsec. Both regions have a similar luminosity, $10^{38.4}$ erg s$^{-1}$.} Hence we are forced to rely upon an estimated characteristic metallicity for the host, based on its absolute magnitude and a radial metallicity gradient, of 12 + log(O/H) = $8.96 \pm 0.04$ dex from the calibration of Boissier \& Prantzos (\citealp{Boi09}). We note that this error does not take into account the intrinsic scatter in the calibration of Boissier \& Prantzos, which is significant. As the determined metallicity is close to the solar value (Asplund et al. \citealp{Asp09}), we have used solar metallicity evolutionary tracks in Section \ref{s2d}.

\subsection{Archival data}

NGC 3389 was observed with the Wide Field and Planetary Camera 2 (WFPC2) on board the \emph{Hubble Space Telescope} (\emph{HST}) on 20 May 2005 ($\sim 1650$ days before explosion). Eight images were obtained in each filter, with total exposure times of 2.81 h in $F606W$ and 2.75 h in $F814W$. The location of SN 2009md fell on the WF2 chip, which has a pixel scale of 0.1\arcsec\ pixel$^{-1}$. Reprocessed data were downloaded from the ESO archive\footnote{http://archive.eso.org}, these data have up to date calibrations applied by the OTF (On-The-Fly) pipeline.

The 3.6-m New Technology Telescope (NTT) observed the site of SN 2009md on 2002 March 25 in the $H$ and $K_\mathrm{s}$ filters with the near-infrared (NIR) camera and spectrograph SOFI, which has a pixel scale of 0.29\arcsec\ pixel$^{-1}$. The $K_\mathrm{s}$ filter data had a total exposure time of 480 s (comprised of 8 separate frames, each with 10 $\times$ 6 s sub-integrations), while the $H$ filter data had a total exposure time of 360 s (8 frames, each with 3 $\times$ 15 s sub-integrations).\footnote{Images from the Subaru Telescope + SuprimeCam and the 3.6-m ESO telescope + EFOSC2 were also examined, but were found to be unusable.}

\subsection{Alignment and photometry}
\label{s2a}

We obtained a $K$-band image of SN 2009md on 2010 Feb 27 (MJD 55254) with NIRI+ALTAIR on the Gemini North telescope as part of our progenitor identification program\footnote{GN-2010A-Q-54.}. We used the f/32 camera on NIRI (Hodapp et al. \citealp{Hod03}) which has a 0.022\arcsec\ pixel scale across a 22.4\arcsec\ $\times$ 22.4\arcsec\ field of view. ALTAIR is the adaptive optics system on Gemini, and provides the $\sim$0.1\arcsec\ resolution needed for precision astrometry. As the SN was too faint to use as a natural guide star for ALTAIR, we used the laser guide star (LGS) to guide for high-order corrections, and the SN itself for tip-tilt corrections.

As the location of the SN is in a crowded field, we obtained off-source images which were then used for sky subtraction of the on-source images. Each image consisted of a 60 s integration (comprised of 2 x 30 s coadds), and after removing bad frames we were left with 1860 s on source. Data were reduced using the {\sc gemini niri} routines within {\sc iraf}\footnote{http://iraf.noao.edu/}, the basic steps consisted of creating a master flat field by median combining the appropriate images, followed by masking of bad pixels and division by the master flat for both on- and off-source images. The sky images taken immediately before and after each series of on-source images were median combined to make a sky image for that sequence. Finally, the sky-subtracted on-source frames were median combined to create the final, reduced image.

After aligning the pre-explosion (WFPC2 $F814W$) and post-explosion (NIRI $K$) images according to the World Coordinate System (WCS), we identified 32 sources common to both images. The coordinates of these sources were measured with the {\sc iraf phot} task using the centroid centering algorithm. The list of matched coordinates was then used as input to the {\sc iraf geomap} task to derive a geometric transformation between the two images, allowing for translation, rotation and independent scaling in the x and y axes. 10 sources were rejected at this stage, as they were outliers of more than one NIRI pixel from the fit. With 22 sources remaining though, we can be confident of not over-fitting the data. The rms error in the fit was 14 mas.

The coordinates of the SN were measured in the NIRI image with the three different centering algorithms offered by the {\sc phot} task: centroid, Gaussian and optimal filtering. The SN is by far the brightest source in the NIRI image, and hence there is no risk of nearby sources within the centering box influencing the centering algorithm. We took the mean of the three centering algorithms as the true (NIRI) pixel coordinates of the SN, with the range (3 mas) as the error. We then transformed these coordinates to WFPC2 pixel coordinates using the previously determined transformation.

\begin{figure*}
\centering
\subfigure[Section of Gemini NIRI post-explosion image, centered on the SN.]{
\includegraphics[height=55mm,width=55mm]{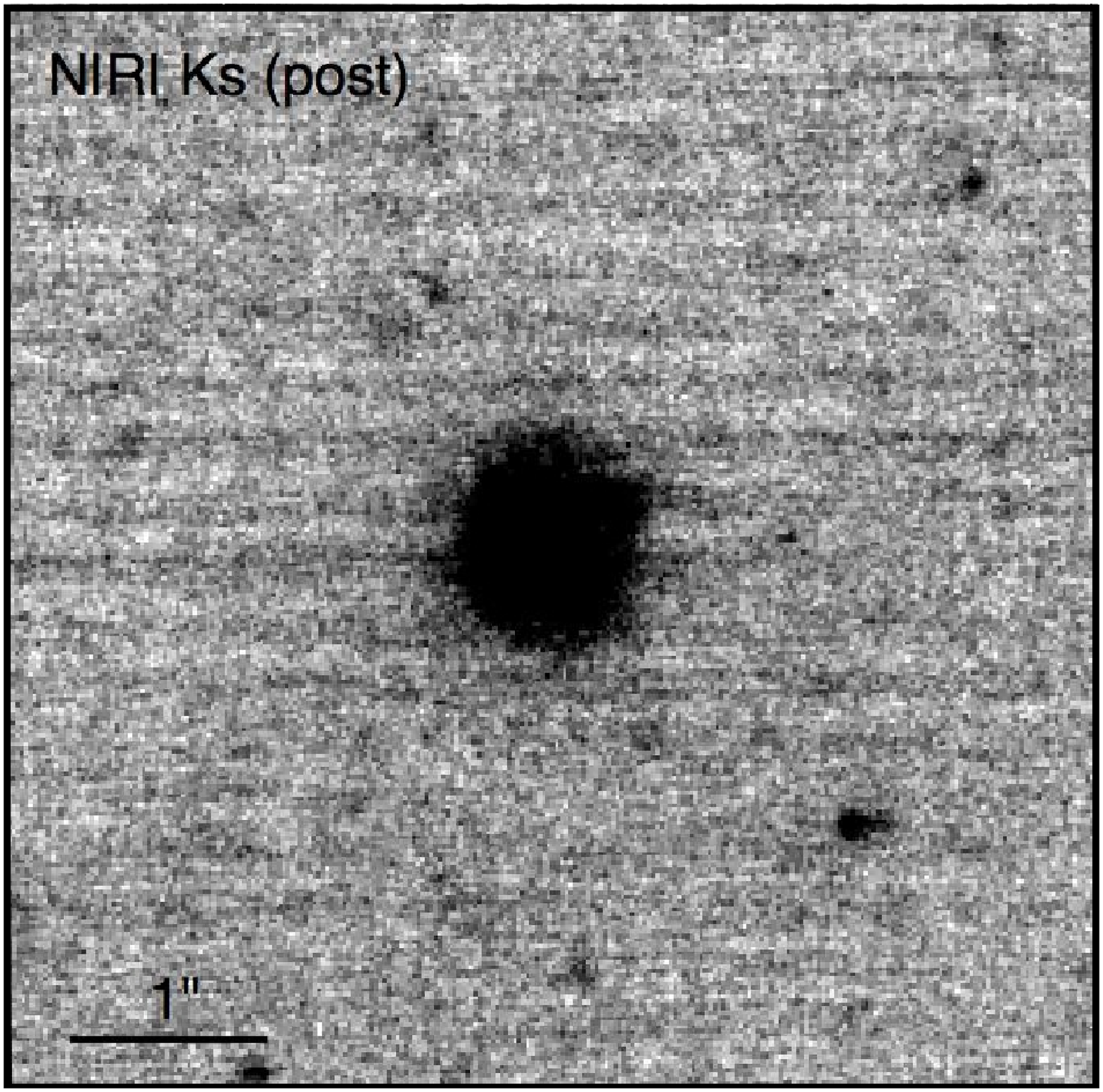}
}
\hspace{2mm}
\subfigure[Section of HST WFPC2 pre-explosion image, centered on the SN location. A blow up of the central region (indicated by a black square is shown in panel (c).]{
\includegraphics[height=55mm,width=55mm]{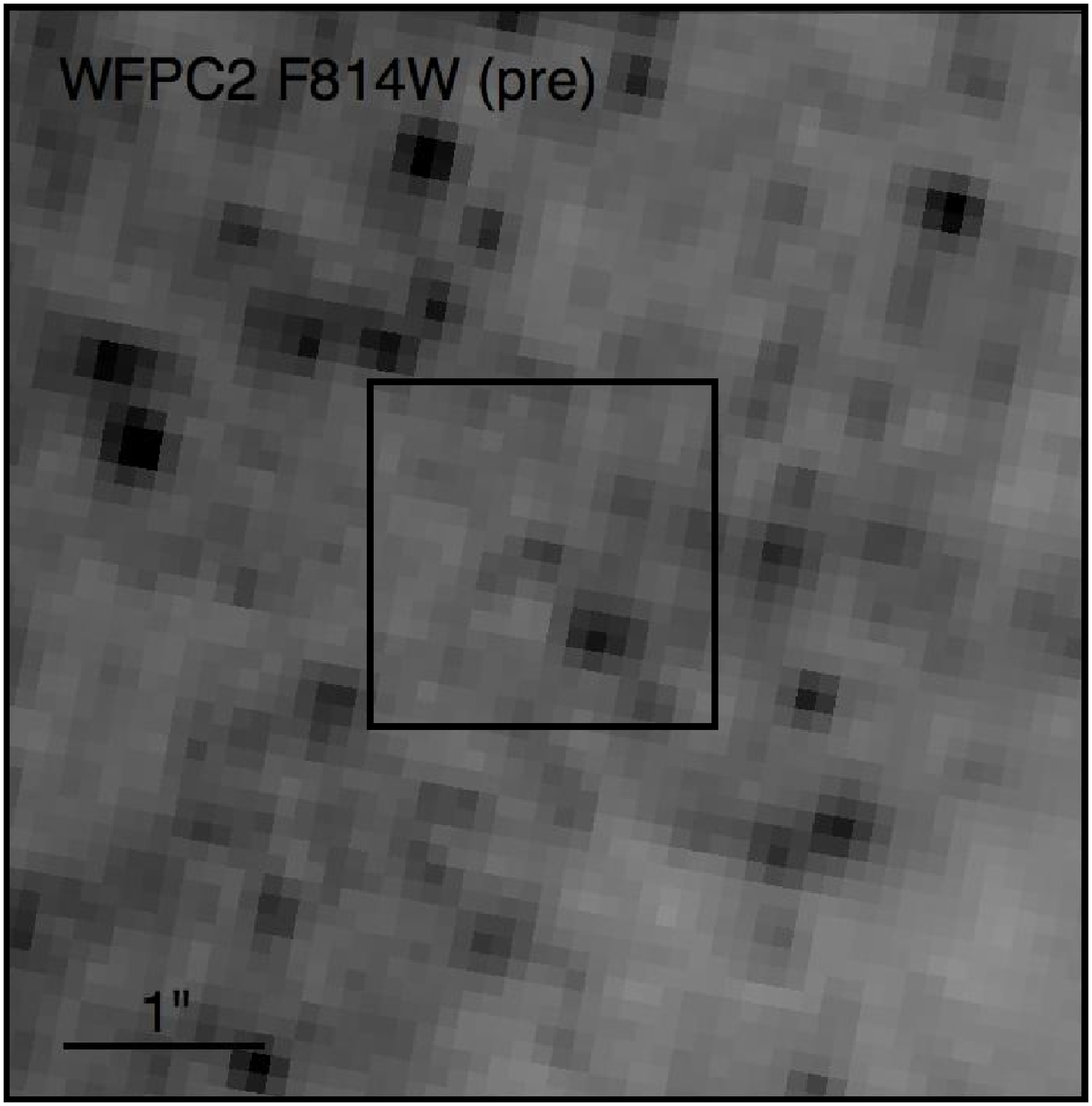}
}
\hspace{2mm}
\subfigure[Blow-up of progenitor (Source A, at intersection of lines) together with nearby sources (B, C and D).]{
\includegraphics[height=55mm,width=55mm]{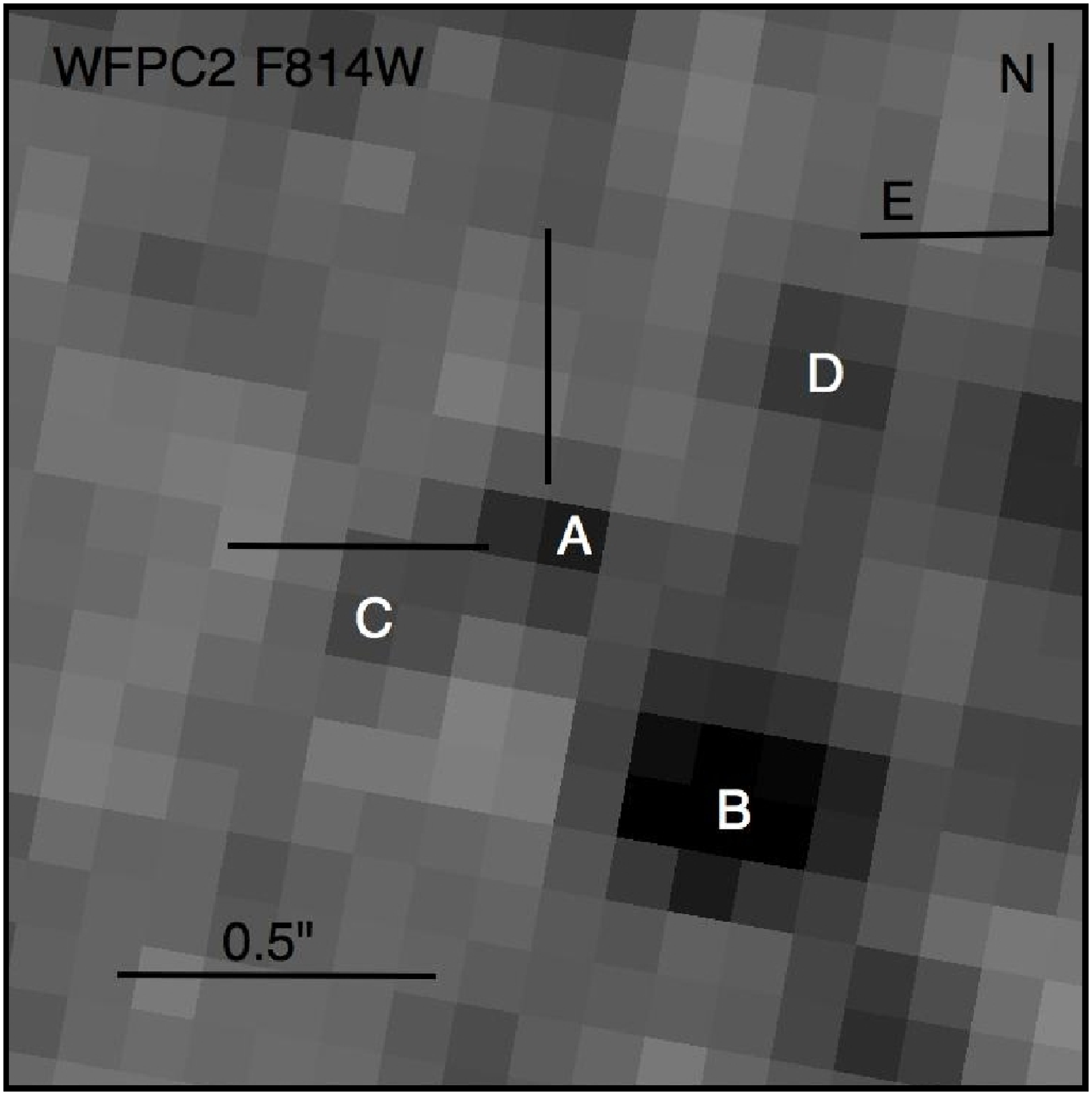}
}
\label{fig-progenitor}
\caption[Optional caption for list of figures]{Pre- and post-explosion images used for progenitor identification}
\label{f3}
\end{figure*}

We can immediately identify a progenitor candidate close to the transformed SN position in the WFPC2 image. Precisely measuring its pixel coordinates is more problematic, as a bright source to the south west will affect the centering algorithms used, introducing a systematic error in all our results. To avoid this, we simultaneously fit a Point Spread Function (PSF) to both the progenitor candidate and all nearby sources, and used the center coordinates as determined by this fitting process as the progenitor coordinates. We used {\sc tiny tim} (Kirst \& Hook \citealp{Kir04}) to model a PSF for the chip and pixel coordinates of the progenitor in the WFPC2 $F814W$ filter image. We then used the {\sc allstar} task within the {\sc iraf daophot} package to fit the PSF to the pre-explosion WFPC2 image. Fitting the progenitor candidate (source A in Fig. \ref{f3}) and the nearby bright source (B), together with two fainter sources (marked C and D) we find the progenitor candidate to be 16 mas from the transformed SN coordinates. Estimating the error in the progenitor coordinates is also difficult as the error introduced by the nearby bright source will affect all three centering algorithms in {\sc phot} similarly. To overcome this, we subtracted the bright nearby source (B) using the model PSF created previously, and measured the position of the progenitor using a 3$\times$3 pixel centering box. While this is a small region to center over, $\sim$ 75 per cent of the flux of our model PSF is within these 9 pixels, and in a region where the background sky is varying on small scale it is necessary to use a small centering box. As before, the range of positions given by the three centering algorithms within {\sc phot} was taken as the error, which in this case was 15 mas.

Adding the uncertainties in progenitor (15 mas) and SN (3 mas) location in quadrature with the rms error in the transformation (14 mas), we find a total error of 21 mas. The progenitor candidate and SN are separated by 16 mas, which is comfortably within the combined fit uncertainty, and so we find the progenitor candidate and SN to be coincident. The progenitor candidate is shown in Fig.~\ref{f3}.

To perform photometry on the identified progenitor candidate, we used the {\sc hstphot} package (Dolphin \citealp{Dol00a}), which is a dedicated package optimized for photometry of under-sampled WFPC2 images. {\sc hstphot} incorporates up-to-date corrections for charge transfer efficiency (CTE), and can convert magnitudes from the $HST$ flight system to the standard $UBVRI$ system (Dolphin \citealp{Dol00b}, \citealp{Dol09}). As {\sc hstphot} cannot be run on drizzled images, we obtained the original individual images in each filter from the MAST archive at STScI. All images were masked for bad pixels based on their associated data quality image; cosmic rays were detected and masked with a median filter.

Using {\sc hstphot}, we found the progenitor magnitude to be  $m_{F606W} = 26.736 \pm 0.15$ mag and $m_{F814W} = 24.895 \pm 0.08$ in the HST flight system, corresponding to a Johnson-Cousins magnitude of $m_{V} = 27.32 \pm 0.15$ mag and $m_{I} = 24.89 \pm 0.08$ mag (using the transformations of Dolphin \citealp{Dol00b}, \citealp{Dol09}). 

As a check on the accuracy of our photometry, we used the {\sc tiny tim} package to create a WFPC2 PSF, appropriate for the chip, progenitor coordinates on the chip and filter for the pipeline drizzled $F814W$ image, which we used to identify our progenitor. PSF-fitting photometry was then simultaneously performed on our progenitor candidate, and on all bright nearby sources using the {\sc iraf daophot} package. The zeropoint for the photometry was taken from the header; we multiplied by PHOTFLAM to obtain a flux, and then used the standard zeropoint for WFPC2 to convert this value to a magnitude. We found a magnitude in the $F814W$ filter for the progenitor which was 0.1 mag brighter than that returned by {\sc hstphot}, which is comparable to our uncertainty.

We also repeated the same procedure for 8 sources, of comparable magnitude to our progenitor candidate, that were detected both by {\sc hstphot} and {\sc daophot}. In all cases {\sc daophot} returned a magnitude that was brighter than {\sc hstphot}. The mean difference in brightness was 0.2 mag, with a standard deviation of 0.1 mag. The reason for this systematic discrepancy is unclear, although we stress that it does not affect the results of this paper in any significant sense, as the discrepancy is comparable to the uncertainty of the output of {\sc hstphot}. We are inclined to favor the values returned by {\sc hstphot} as the package is optimized specifically for undersampled WFPC2 images. As a more qualitative indicator of the accuracy of our photometry, the supergiant sequence shown in Fig.~\ref{f2} appears reasonable, in that we do not see any gross discrepancies with the standard sequence that would indicate there is a systematic error in our photometry.

While the SN and progenitor candidate are spatially coincident to within the errors, in deep, crowded fields in such a nearby galaxy we must be wary of a chance alignment between the SN and an unrelated source. We will argue against this being the case using several independent lines of reasoning. Firstly, we consider the likelihood of a chance alignment based simply on the number of detected sources in the field. While we find a total error of $\sim$21 mas in the alignment for SN 2009md, even if we found a source coincident to within a $\sim$40 mas error in alignment, we would likely consider this a plausible progenitor candidate. Hence we took a region around the SN covering 36 arcsec$^2$, and used the number of sources detected by {\sc hstphot} at a significance of $\geq$5$\sigma$ (428 objects), together with an arbitrary ``association'' distance of 40 mas around each source to calculate a probability of a chance alignment, which we find to be 6 per cent. This is a very rough estimate, as some of the regions around each source will overlap, reducing this probability. Of course, if we found a coincident source at the 4$\sigma$ level, or a source 41 mas from the SN, we would likely consider it an interesting association, so the values adopted in this calculation are somewhat arbitrary. Nonetheless, 6 per cent is probably a sensible estimate of a chance alignment for any source with the SN. We can further reduce this probability by considering the likelihood of finding a source coincident with the SN, and with a colour redder than some set value. If we repeat the calculation, but only for sources with a $V-I$ colour greater than 1.25 mag, corresponding to a supergiant of spectral type K0 or later, we find a likelihood of a chance alignment of $<$ 2 per cent. 

\subsection{SOFI data}
\label{s2b}

\begin{figure}
\includegraphics[width=0.48\textwidth,angle=0]{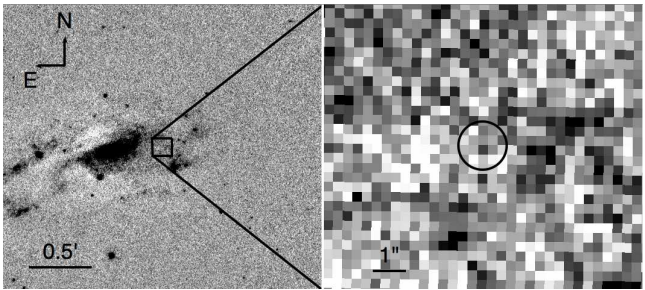}
\caption{Subsections of pre-explosion NTT SOFI image of NGC 3389, taken with the Ks filter. Right panel shows a magnified portion of the field (indicated by the box). The field is centered on the SN position, as discussed in the text. The circle corresponds to 5 times the positional uncertainty in the transformation and SN coordinates.} 
\label{f4}
\end{figure}

The SOFI data were downloaded from the ESO archive, together with appropriate calibration frames (flat fields and dark frames) taken on the same night. The data were reduced using standard {\sc iraf} tasks. Subtraction of the flux from the sky is crucial in the NIR, unfortunately in this case there were no off-source sky-frames, and the dithers between the on-source frames were too small to allow a clean sky image to be constructed. We have created a median combined sky image using the on-source science frames, and subtracted this from each individual frame using the {\sc iraf xdimsum} package. The individual reduced science frames were then aligned and median combined to produce a final, reduced image, which is shown in Fig. \ref{f4}.

We registered the $H$ and $K_\mathrm{s}$ pre-explosion images to the post-explosion NTT+EFOSC2 $R$-band image from 2010 January 22, as the field of view covered by the post-explosion NIRI image was too small to identify sufficient numbers of common sources. The same basic procedure was followed as when aligning the WFPC2 and NIRI images, 15 sources common to both images were identified. The positions of these reference stars were measured in both frames with {\sc iraf phot}. The list of matched coordinates were then used with {\sc iraf geomap} to derive a transformation, allowing for translation, rotation and a scaling factor. Two sources were rejected as outliers from the fit, the final rms error in the transformation was 0.541 pixels, which corresponds to 130 mas for the 0.24\arcsec\ binned pixels of EFOSC2. As before, we measured the position of the SN in the EFOSC2 image using the three centering algorithms within {\sc phot}, and find a range of 4 mas between the measurements. Using the {\sc geoxytran} and the derived transformation, we transformed the measured coordinates of the SN to the pre-explosion SOFI image.

There is no obvious source present at the location of SN 2009md  in either the $H$- or $K_\mathrm{s}$-band pre-explosion image. A non-detection is still of interest, as we can use it to constrain the magnitude of the progenitor in the near-infrared. A photometric zeropoint of $ZP_\mathrm{Ks}=23.9$ mag was determined for the $K_\mathrm{s}$ image using PSF-fitting photometry with reference to catalogued Two Micron All-Sky Survey (2MASS) sources in the field. As the main source of noise in the near-infrared is from the bright sky background, this is the dominant factor in our limiting magnitude. We measured the standard deviation of the sky background at the SN position using the {\sc iraf imexam} task, and found it to be $\sim 3$ ADU. Assuming that the central pixel of a detected source at confidence level of 3$\sigma$  will have a flux at center of 3 times the standard deviation, we calculated the magnitude of a PSF with a central pixel 3$\sigma$ above the background, at the location of the SN (and using the same aperture size as used for the sequence stars). We find a limiting magnitude of $K_\mathrm{s} > 19.4$ mag. The $H$ filter image did not yield a useful stringent limit.

\subsection{Extinction estimates}
\label{s2c}

The foreground extinction towards NGC 3389 was taken, via NED\footnote{http://nedwww.ipac.caltech.edu/}, from the Schlegel, Finkbeiner \& Davis (\citealp{Sch98}) dust maps, which give a colour excess $E(B-V) = 0.027$ mag. Estimating the total line-of-sight extinction to the SN is more difficult as there is not a single technique which gives reliable results. Turatto, Benetti \& Cappellaro (\citealp{Tur03}) give an empirical relation between the equivalent width of the Na {\sc i} doublet, EW(Na {\sc i} D), at 5890 and 5896~\AA\ and the colour excess $E(B-V)$. The Na {\sc i} D observed in the spectrum of SN 2009md is a blend of the contribution from the host galaxy and the Milky Way, so in this case we are measuring  the \emph{total} line of sight extinction. We measured the strength of the Na {\sc i} doublet using a combination of the two spectra obtained on Dec 7 and Dec 10 (Section \ref{s3c}) to improve the signal to noise ratio, to find EW(Na{ \sc i} D) = 1.32 \AA. Applying the relation of Turatto et al. gives $E(B-V)=0.21$ mag. However, there is considerable scatter in the relation of Turatto et al., typically on the order of $\pm 0.05$ mag for EW(Na{ \sc i} D) $\sim 1$ \AA. We can also measure the position of the center of the Na{ \sc i} D absorption, which we find to be at 5907 \AA. As this is approximately midway between the central wavelengths of this doublet for the Milky Way (rest) and host galaxy velocities, it seems likely that the internal extinction in the host is comparable to that in the foreground. From this, we would expect a total colour excess $E(B-V) = 0.05$ mag.

\begin{figure}
\includegraphics[width=0.45\textwidth,angle=270]{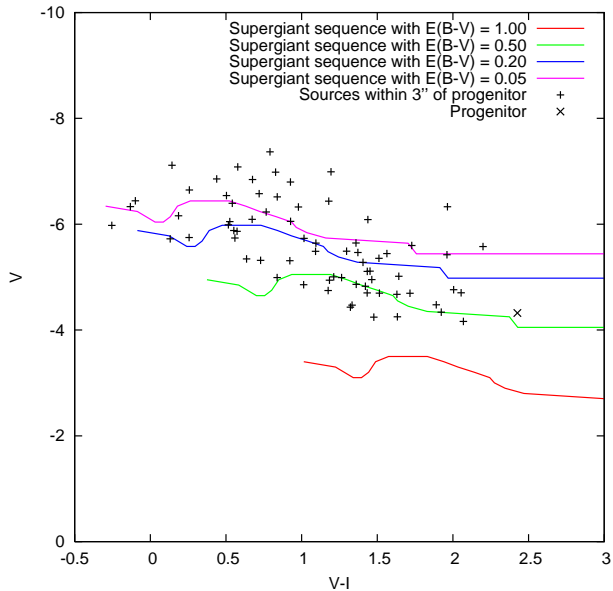}
\caption{The absolute magnitude of all sources detected within 3\arcsec\ of SN 2009md, detected in archival \emph{HST} WFPC2 $F555W$ and $F814W$ images at $\geq$5$\sigma$, after applying sharpness and $\chi^2$ cuts as discussed in text. The progenitor is marked with an x, nearby sources with a plus symbol; the progenitor appears to be very red, which supports its status as an evolved object. The coloured lines are a standard sequence of supergiants from Drilling \& Landolt (\citealp{Dri00}), converted to Johnson-Cousins as per Bessell (\citealp{Bes79}), with different extinction values applied. The presence of blue sources seen close to the SN support low levels of extinction; the best match appears to be for a colour excess of $E(B-V) \sim 0.05-0.2$ mag.}
\label{f2}
\end{figure}

More qualitative measures of extinction involve comparing the colour and magnitude of the SN to similar objects; a comparison of the colour evolution of SN 2009md during the photospheric phase (Fig. \ref{f8}), corrected for a colour excess $E(B-V)=0.05$ mag, is a good match for the Type IIP SNe 1999em and 2005cs. We can also compare the colours and absolute magnitudes of nearby sources in pre-explosion images against the standard supergiant sequence of Drilling \& Landolt (\citealp{Dri00}). We have taken the magnitudes of all sources found in HST WFPC2 pre-explosion images, within an arbitrary 3\arcsec\ radius of the progenitor, as discussed in Section \ref{s2a}, and plotted on a Hertzprung-Russell diagram in Fig. \ref{f2}. We have also applied the selection criteria that sources must be found in both the $F606W$ and $F814W$ filters at a significance $\geq$5 $\sigma$, and must have $\chi^2$ and sharpness statistics that are consistent with a non-extended source. The magnitudes of the sources in Fig. \ref{f2} have not been corrected for reddening, only for the distance modulus. In addition, we plot a supergiant sequence from Drilling \& Landolt with four differing amounts of reddening. As can be seen, the surrounding sources do not appear to be heavily reddened, the best match being with $E(B-V) = 0.2$ mag. It is important to remember however, that at the distance of NGC 3389, 3\arcsec\ corresponds to $\sim 300$ pc, and reddening can vary over such large scales.

Weaknesses can be identified in all the techniques used for estimating the reddening; low resolution spectra make it difficult to measure equivalent widths and line centers precisely, doubts have also been cast on the degree to which extinction and Na {\sc i}  correlate (Poznanski et al. \citealp{Poz11}). Comparison of the local supergiant sequence is a crude test, at the distance to NGC 3389. Comparison of the colours of SN 2009md to those of other SNe is based on the assumption that these SNe are intrinsically similar. These caveats notwithstanding, we will adopt a value for $E(B-V) = 0.1^{+0.1}_{-0.05}$ mag. We also note that Fig. \ref{f2} provides support for the adopted distance to NGC 3389. From the arguments presented above, the reddening is not particularly high, and so in this case the fact that the identified sources do not appear to be systematically brighter or fainter than the supergiant sequence indicates that the distance we have adopted is correct.

\subsection{Progenitor analysis: Luminosity and mass estimates}
\label{s2d}

In Smartt et al. (\citealp{Sma09b}) the luminosities and temperatures of SN progenitors were calculated using \emph{observationally} derived supergiant colours and bolometric corrections from the compilation of Drilling \& Landolt (\citealp{Dri00}). However, the references from which Drilling \& Landolt compiled their data are now over forty years old, and so a re-evaluation and update of this methodology appears timely. In particular, Drilling \& Landolt take their supergiant colours from Johnson (\citealp{Joh66}), where the magnitudes are in the photometric system of Johnson (\citealp{Joh64}). The $I$ band defined by Johnson in these papers is substantially different from the modern Cousins $I$ band which is used in the Johnson-Cousins system. Hence all $V-I$ colours in Drilling \& Landolt (\citealp{Dri00}) should be converted from Johnson to Johnson-Cousins before use. A suitable transformation is given by Bessell (\citealp{Bes79}), which can be trivially applied to Drilling \& Landolt.

However, we have chosen instead to use synthetic photometry of model spectra to define our bolometric corrections and colours, and hence derive a temperature and luminosity for the progenitor candidate (see, for example, Eldridge, Mattila \& Smartt \citealp{Eld07}). Our reasons for doing so are several: by using synthetic photometry we can work in the $HST$ filter system rather than having to convert to Johnson-Cousins filters, and we are not at risk of introducing errors due to the uncertain observed extinction towards red supergiants from the literature.

We have taken model spectra of massive red supergiants produced with the MARCS code (Gustafsson et al. \citealp{Gus08}) for a range of temperatures, and used the {\sc synphot} package within {\sc iraf} to produce synthetic colours in the \emph{HST} filter system, taking appropriate zeropoints from Dolphin (\citealp{Dol09}). We have listed the derived \emph{HST} filter system colours, together with the bolometric corrections from Levesque (\citealp{Lev05}) for a range of model spectra at different temperatures in Table \ref{t_marcs}.

Taking the values for the progenitor magnitude from Section \ref{s2a} together with the extinction and distance modulus as discussed in Sections \ref{s2f} and \ref{s2c}, we calculate the progenitor absolute magnitude in the \emph{HST} filter system to be M$_{F606W}=-5.18 ^{+0.38}_{-0.29}$, and the $F606W-F814W$ colour to be $1.74\pm 0.20$. Using the values in Table \ref{t_marcs} to determine the bolometric correction and temperature for this colour, we find a bolometric correction of $-2.01\pm0.3$ mag for the progenitor of SN 2009md, and an effective temperature of 3530$^{+70}_{-40}$ K, which according to Levesque (\citealp{Lev05}) is consistent with a star of spectral type M4. Calculating the $F606W-V$ colour from Table \ref{t_marcs} (which is essentially flat at $-0.59$ mag), and using this with the bolometric correction, we find a bolometric magnitude for the progenitor of $-6.60^{+0.43}_{-0.49}$ mag. We use the standard relation for luminosity and bolometric magnitude

\begin{equation}
\text{log} L/L_{\odot} = {{M_ \mathrm{bol} - 4.74}\over{-2.5}}
\label{eq_lum}
\end{equation}

\begin{table*}
\caption{Synthetic colours in the WFPC2 flight system from MARCS models and {\sc synphot}.}
\begin{center}
\begin{tabular}{lccccc}
\hline
T$_{eff}$ 		 	& $F450W-F555W$		& $F555W-F606W$	& $F606W-F814W$ 	& $F555W-V$	& BC \\
\hline
3400				& 1.288				& 0.633			& 2.307			& 0.044		& -2.810\\
3500				& 1.416				& 0.616			& 1.852			& 0.029	 	& -2.180\\
3600				& 1.475				& 0.602			& 1.543			& 0.019		& -1.750\\
3700				& 1.476				& 0.583			& 1.341			& 0.015		& -1.450\\
3800				& 1.442				& 0.560			& 1.203			& 0.016		& -1.230\\
3900				& 1.391				& 0.536			& 1.105			& 0.019		& -1.060\\
4000				& 1.336				& 0.511			& 1.030			& 0.024		& -0.920\\
\hline
\end{tabular}	 				
\end{center}
\label{t_marcs}
\end{table*}

to find a progenitor luminosity of log $L$/\lsun = 4.54 $\pm0.19$ dex.

\begin{figure}
\includegraphics[width=0.48\textwidth,angle=0]{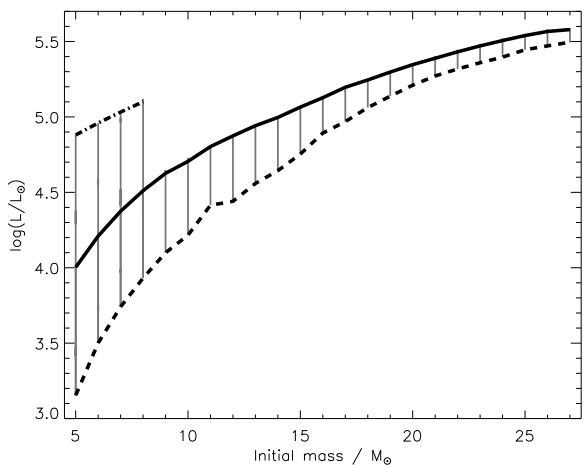}
\caption{Luminosity as a function of mass from the STARS models. The solid line is the luminosity at model endpoint, or for models lower than 8 \msun, the pre second dredge-up luminosity. The dot-dashed line is the progenitor luminosity after second dredge-up. The dashed line is the helium core luminosity of the model, from which the upper mass limit is obtained.} 
\label{f5}
\end{figure}

To use this luminosity to estimate a mass for the progenitor, it is necessary to compare to a series of evolutionary tracks. As discussed in Section \ref{s4b}, we have used the STARS models (Eldridge \& Tout \citealp{Eld04}) for comparison in this work. We find a best match for a pre-second dredge-up model of $8.5^{+6.5}_{-1.5}$ \msun, as shown in Fig. \ref{f5}, although we stress that in this mass regime there are problems with stellar evolutionary models at later burning stages, which we will return to in Section \ref{s4b}. The upper limit for the mass is taken from the mass of the progenitor that would have a luminosity of 4.73 dex at the end of core He burning (for the reasons discussed in Smartt et al. \citealp{Sma09b}), which corresponds to 15 \msun, while the lower limit is found to be 7 \msun\ from the uncertainty in the progenitor luminosity when plotted on the pre-second dredge-up track. The masses quoted here and in the rest of the paper refer to the zero-age main sequence mass (ZAMS) of the progenitor, although for stars in this regime, mass loss over the lifetime of the star is quite low (0.6 \msun\ for a 9 \msun\ STARS model at solar metallicity). 

For the non-detection in SOFI images, we have taken the bolometric corrections for the $K_\mathrm{s}$ filter from Levesque et al. (\citealp{Lev05}), these range from 3.16 to 2.33 mag for effective temperatures between $\sim$3200 K (early M type) and $\sim$4300 K (early K type) respectively. If we assume the effective temperature measured from the WFPC2 colours, the bolometric correction is closer to the former than the latter, however we will consider the two possibilities separately. For a cooler progenitor (and hence a larger bolometric correction), we find that the limiting luminosity log$L$ is $5.5\pm0.1$ \lsun, with a corresponding upper mass limit of $M < 27$ \msun. For the hotter progenitor, we find a mass limit of $M< 36$ \msun. We have followed the maxim that it is better to err on the side of caution when presenting limits from non-detection, and so our final upper mass limits perhaps err on the conservative side.%

\subsection{Ruling out an SAGB progenitor and internal dust extinction}
\label{s2e}

One factor that we have neglected up to this point is the possibility of circumstellar dust that is destroyed in the SN explosion. Several authors (for example Dwek \citealp{Dwe83}; Waxman \& Draine \citealp{Wax00}) have suggested that the initial X-ray and UV flash of a SN or gamma-ray burst could photoevaporate large quantities of dust in the vicinity of the progenitor. We would see no sign of this photoevaporated dust in the SN spectrum, and so there exists a possibility that we have underestimated the extinction towards the SN progenitor. Furthermore, red supergiants in the Galaxy and Magellanic Clouds are observed to suffer excess extinction when compared to nearby OB stars (Levesque et al. \citealp{Lev05}; Massey et al. \citealp{Mas05}), which has been interpreted as evidence for high levels of dust around these stars. Indeed, taking the sample of red supergiants from Levesque et al. (\citealp{Lev05}), we find a mean $A_V$ of 0.7 mag. One caveat which must be attached to this however is that the extinction towards red supergiants will likely vary as a function of metallicity. Stars at a lower metallicity have a lower mass loss rate, and so we may expect to find less circumstellar dust, and hence extinction. However, as we have adopted a solar metallicity (see Section \ref{s2f}), the comparison with Levesque et al. (\citealp{Lev05}) is appropriate.

\begin{figure}
\includegraphics[width=0.35\textwidth,angle=270]{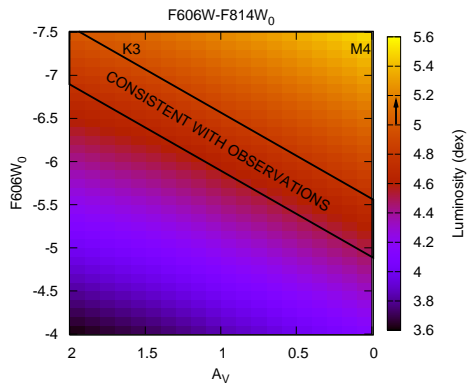}
\caption{Progenitor luminosity as a function of extinction in the $V$-band and intrinsic $F606W$-filter absolute magnitude. On the lower x axis is extinction in $V$, while the upper x axis gives the corresponding intrinsic $F606W-F814W$ colour a progenitor would have with this degree of extinction, given that it has an apparent colour of $F606W-F814W$ = 1.84 mag. On the y axis is plotted the intrinsic progenitor magnitude in the $F606W$ filter. The bolometric luminosity, as indicated by the colour bar, has been calculated with Eq. \ref{eq_lum}, using appropriate bolometric corrections (from Levesque et al. \citealp{Lev05}). Combinations of $A_V$ and $F606W$ which are consistent with observations are marked. The black arrow on the colour bar indicates the typical SAGB luminosity that is expected to give rise to an electron-capture SN (Poelarends et al. \citealp{Poe08}).} 
\label{f23}
\end{figure}

To consider the impact of dust on our progenitor analysis, we recalculated the range of intrinsic progenitor colours and luminosities which are consistent with the observed progenitor $F606W$ filter magnitude and $F606W-F814W$ colour, leaving the extinction as a free parameter, as shown in Fig. \ref{f23}. We have used the bolometric corrections for supergiants from Levesque et al., these are likely appropriate for SAGB stars as well, as they are chiefly a function of colour (eg. Drilling \& Landolt \citealp{Dri00}). We find that we can only produce a progenitor luminosity consistent with an SAGB star with $\gtrsim$ 2 magnitudes of extinction in the $V$-band. This level of reddening would mean that the progenitor is actually an early K supergiant, which is too hot to be an SAGB star, and so we can exclude an SAGB star as the progenitor for SN 2009md. In fact, it is quite difficult to make the progenitor more luminous by invoking dust - as the extinction is increased, the progenitor becomes intrinsically brighter in $F606W$, however to be self-consistent it must also be hotter, which implies a smaller bolometric correction. The two factors largely cancel each other out, leaving a progenitor of only marginally higher luminosity.

As can be see in Fig. \ref{f23}, we cannot exclude a higher mass, and hence more luminous, early K type red supergiant progenitor. However, we consider such a scenario unlikely. Taking the sample of red supergiants from Levesque et al. (\citealp{Lev05}), and selecting only those stars with both an observed $V$ magnitude and a $V-I$ colour which are within 1 mag of that of the progenitor of SN 2009md, we find a mean $A_V$ of 0.38 mag, with an rms scatter of 0.58 mag. Furthermore, only 12 per cent of the stars in this subset have a value of $A_V >$ 1 mag, and only 4 per cent have $A_V >$ 1.5 mag. We also note that, as expected, the heavily extinguished stars in the Levesque sample tend to have a redder colour than our progenitor candidate. An increase in $A_V$ of $\sim$1 mag corresponds to an increase of  $\sim$ 10 per cent in the initial mass as calculated in Section 2.6. This is smaller than our error bars and we thus conclude that it is unlikely that our results are affected by excess extinction.

\section{Follow-up observations}
\label{s3}

\subsection{Photometric follow-up}

\label{s3a}

A campaign of follow-up observations\footnote{This is paper is based on ESO-NTT and TNG long term programs, in the framework of a large international collaboration for supernova research. For the composition of the Collaboration and its scientific goals we refer the reader to our web pages (http://graspa.oapd.inaf.it/).} was initiated for SN 2009md shortly after discovery, with data obtained from the Liverpool Telescope (LT) + RATCam and SupIRCam, the Faulkes Telescope North (FTN) + EM01, the New Technology Telescope (NTT) + EFOSC2 and SOFI, the Nordic Optical Telescope (NOT) + ALFOSC, the Telescopio Nazionale Galileo (TNG) + NICS, the Calar Alto 2.2-m telescope + CAFOS,  and the Wise Observatory Telescope + PI and LAIWO. The LT and FTN are identical 2-m robotic telescopes, located at  Roque de los Muchachos, La Palma and Haleakala Observatory, Hawaii respectively. Both telescopes have a standard pipeline that supplies reduced and calibrated images. For reduction of all other optical data, the {\sc quba}\footnote{{\sc IRAF} based Python package for photometry and spectroscopy developed by SV (Queens University Belfast). See Valenti et al. \citealp{Val11} for details.} pipeline was used. The NTT + SOFI NIR data were reduced with the {\sc sofi}\footnote{ftp://ftp.eso.org/pub/dfs/pipelines/sofi} pipeline and the TNG + NICS Near-Infrared (NIR) data in {\sc IRAF} using standard NIR reduction techniques.

All photometry was performed with the {\sc quba} pipeline. Photometric zeropoints and colour terms in the optical bands were determined for individual nights with aperture photometry of Landolt fields (Landolt \citealp{Lan92}). Optical magnitudes for a sequence of stars in the SN field (Fig. \ref{f1}) were then measured with PSF-fitting photometry. The final magnitudes of the local sequence stars were obtained averaging the measurements in 5 photometric 
nights. The coordinates and the average optical magnitudes of these stars are reported in Table \ref{t_seq}. Corrected zeropoints and colour terms for non-photometric nights were determined with reference to the average magnitudes in Table \ref{t_phot}. NIR magnitudes for the local sequence stars were taken from the 2 Micron All Sky Survey (2MASS) catalogue (Skrutskie et al. \citealp{Skr06}). The magnitudes of the SN at each epoch in the optical and NIR were measured with PSF-fitting, and are listed in Tables  \ref{t_phot} and  \ref{t_nirphot}.

In this and the following sections, we will chiefly compare SN 2009md to the sub-luminous Type IIP SN 2005cs (Pastorello et al. \citealp{Pas09}). Our choice of comparison is motivated by the availability of photometry and spectra for this object, the detailed analysis in the literature, and most importantly, the fact that a progenitor was also identified for SN 2005cs. We will also make comparisons to SN 1999em as an example of a canonical Type IIP SN, and in some cases to the peculiar Type II SN 1987A as this is the best monitored and understood SN. Distances and reddening for SNe 2005cs, 1999em and 1987A were taken from Pastorello et al. (\citealp{Pas09}), Smartt et al. (\citealp{Sma09b}) and Suntzeff \& Bouchet (\citealp{Sun90}) respectively. The extinction as a function of wavelength for these SNe was calculated using the extinction law of Cardelli, Clayton \& Mathis (\citealp{Car89}).

\begin{figure}
\includegraphics[width=0.48\textwidth,angle=0]{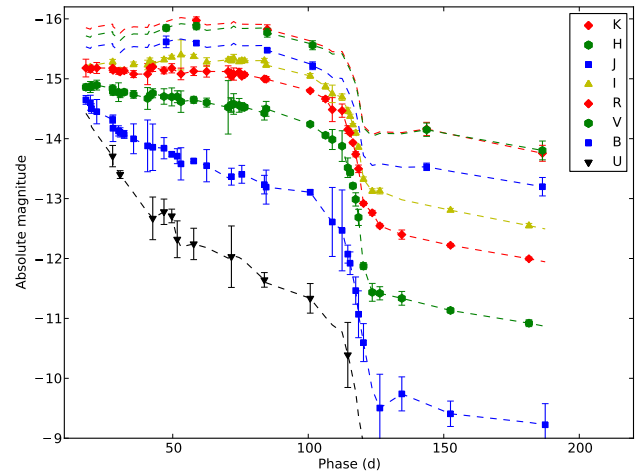}
\caption{Full evolution of SN 2009md in UBVRIJHK filters. Solid points are the measured data points, dashed lines show the interpolated magnitudes used to compute the bolometric lightcurve.}
\label{f6}
\end{figure}

Figure \ref{f6} shows the evolution of SN 2009md in absolute magnitudes, the SN is clearly seen to be a Type IIP SN. The plateau phase of Type IIP SNe is governed by the release of the thermal explosion energy, and for SN 2009md lasts for $\sim$120 d. The magnitude of the SN is roughly constant during the plateau phase in the $I$-band, but shows progressively larger rates of decline in bluer filters due to cooling and increased line blanketing. At the end of the plateau, when the thermal energy source created by the explosion is exhausted, the SN fades by $\sim$2.5 mag in $\sim$1 week. The tail phase that follows is governed by heating from the radioactive decay of $^{56}$Co (the decay product of $^{56}$Ni), making it possible to determine the ejected mass of $^{56}$Ni. We will return to this issue in Section \ref{s3b}.

\begin{figure}
\includegraphics[width=0.52\textwidth,angle=0]{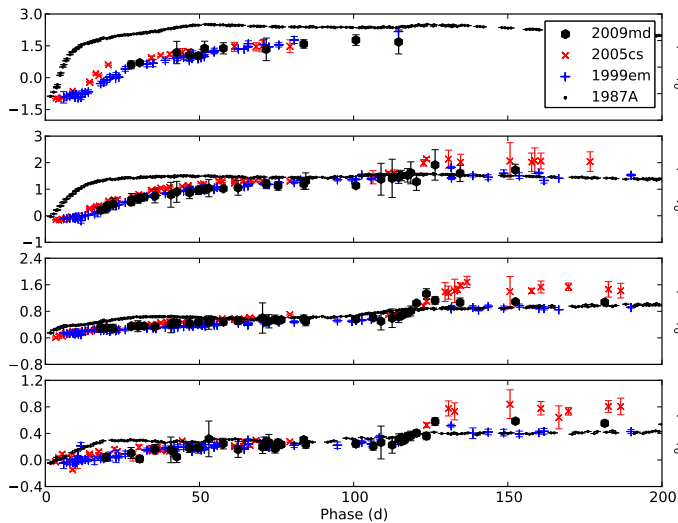}
\caption{Colour evolution of SN 2009md, as compared to SNe 2005cs, 1999em and 1987A. The reddening at the end of the plateau for the sub-luminous Type IIP SNe is evident.}
\label{f8}
\end{figure}

Figure \ref{f8} shows the colour evolution of SN 2009md, as compared to SNe 1999em, 2005cs, and 1987A. During the plateau phase we see a common colour evolution among the Type IIP SNe until the end of the plateau at $\sim$120 d. Beyond this phase, the colour evolution of all Type IIP SNe show a trend towards redder colours due to the change in the temperature as the SN drops off the plateau. The colour change for SNe 2009md and 2005cs in this phase is more pronounced, as previously noted for sub-luminous Type IIP SNe (e.g. fig. 4 in Maguire et al. \citealp{Mag10a}).

\begin{table*}
\caption{Magnitudes of local sequence stars used to calibrate the photometry of SN 2009md. Typical errors in magnitudes for sequence stars are $\lesssim 0.05$ mag.}
\begin{center}
\begin{tabular}{lccccccc}
\hline
Identifier &RA        & Dec   &$U$  &$B$             	&$V$			&$R$              	& $I$	\\
\hline
1 & $10^h~48^m~28.1^s$ & $12\deg~32\arcmin~25\arcsec$ & 19.544 & 18.573 & 17.677 & 17.043 & 16.579 \\
2 & $10^h~48^m~27.9^s$ & $12\deg~31\arcmin~09\arcsec$ & 16.760 & 16.516 & 15.710 & 15.191 & 14.698 \\
3 & $10^h~48^m~33.7^s$ & $12\deg~31\arcmin~35\arcsec$ & 20.621 & 20.621 & 19.340 & 18.351 & 17.313 \\
4 & $10^h~48^m~19.6^s$ & $12\deg~30\arcmin~45\arcsec$ & 20.392 & 19.032 & 17.748 & 16.867 & 16.148 \\
5 & $10^h~48^m~21.6^s$ & $12\deg~33\arcmin~47\arcsec$ & 17.536 & 17.341 & 16.757 & 16.351 & 16.028 \\
6 & $10^h~48^m~31.6^s$ & $12\deg~34\arcmin~19\arcsec$ & 15.156 & 14.922 & 14.283 & 13.866 & 13.552 \\
\hline
\end{tabular}	 				
\end{center}
\label{t_seq}
\end{table*}

\begin{table*}
\caption{Log of optical photometric measurements of SN 2009md. Errors are given in parentheses.}
\begin{center}
\begin{tabular}{lccccccr}
\hline
JD-2400000.5 &Phase (d) &$U$ &$B$ &$V$	&$R$ &$I$ &Instrument \\
\hline
55179.96 & 17.96 &   &17.40 (0.07) &17.09 (0.05) &16.72 (0.15) &  & FTN (EM01) \\
55181.61 & 19.61 &   &17.45 (0.12) &17.09 (0.09) &16.74 (0.05) &16.64 (0.03) & WISE (PI-CCD) \\
55182.02 & 20.02 &   &17.54 (0.08) &17.07 (0.07) &16.72 (0.03) &  & FTN (EM01) \\
55183.96 & 21.96 &   &17.60 (0.20) &17.05 (0.08) &16.72 (0.09) &  & FTN (EM01) \\
55189.83 & 27.83 & 18.42 (0.18) &17.73 (0.06) &17.11 (0.06) &16.72 (0.07) &16.55 (0.04) & NTT (EFOSC) \\
55189.96 & 27.96 &   &17.87 (0.22) &17.16 (0.06) &16.76 (0.04) &  & FTN (EM01) \\
55192.00 & 30.00 &   &17.92 (0.09) &17.17 (0.16) &16.77 (0.05) &  & FTN (EM01) \\
55192.60 & 30.60 & 18.73 (0.07) &17.94 (0.08) &17.20 (0.07) &16.78 (0.03) &16.70 (0.04) & LT (RATCam) \\
55194.09 & 32.09 &   &17.98 (0.07) &17.17 (0.05) &16.76 (0.04) &  & FTN (EM01) \\
55197.55 & 35.55 &   &18.05 (0.25) &17.21 (0.06) &16.82 (0.06) &16.59 (0.04) & LT (RATCam) \\
55202.72 & 40.72 &   &18.17 (0.43) &17.28 (0.18) &16.82 (0.18) &16.60 (0.08) & CALTO (CAFOS) \\
55203.57 & 41.57 &   &  &17.23 (0.05) &16.72 (0.06) &16.53 (0.04) & WISE (LAIWO) \\
55204.63 & 42.63 & 19.46 (0.36) &18.19 (0.39) &17.19 (0.07) &16.70 (0.08) &16.59 (0.04) & LT (RATCam) \\
55208.77 & 46.77 & 19.35 (0.21) &18.21 (0.18) &17.24 (0.13) &16.76 (0.06) &16.51 (0.04) & LT (RATCam) \\
55211.64 & 49.64 & 19.42 (0.11) &18.31 (0.04) &17.25 (0.15) &16.72 (0.08) &16.47 (0.03) & LT (RATCam) \\
55213.66 & 51.66 & 19.81 (0.31) &18.34 (0.12) &17.25 (0.10) &  &  & LT (RATCam) \\
55215.03 & 53.03 &   &18.47 (0.27) &17.33 (0.18) &16.82 (0.10) &16.43 (0.25) & Mt.Ekar 1.82m (AFOSC) \\
55219.74 & 57.74 & 19.89 (0.26) &18.42 (0.03) &17.30 (0.06) &16.77 (0.07) &16.46 (0.04) & NTT (EFOSC) \\
55224.51 & 62.51 &   &18.50 (0.27) &17.34 (0.07) &16.78 (0.09) &16.55 (0.07) & LT (RATCam) \\
55232.45 & 70.45 &   &  &17.42 (0.45) &16.78 (0.10) &16.51 (0.03) & LT (RATCam) \\
55233.47 & 71.47 &   &  &17.39 (0.12) &16.80 (0.04) &16.52 (0.02) & LT (RATCam) \\
55233.65 & 71.65 & 20.10 (0.51) &18.68 (0.14) &17.39 (0.10) &16.86 (0.04) &16.51 (0.05) & NOT (ALFOSC) \\
55234.49 & 72.49 &   &  &17.37 (0.17) &16.82 (0.10) &16.49 (0.06) & LT (RATCam) \\
55236.50 & 74.50 &   &  &17.39 (0.12) &16.79 (0.06) &16.55 (0.02) & LT (RATCam) \\
55237.48 & 75.48 &   &18.64 (0.15) &17.42 (0.07) &16.86 (0.03) &16.53 (0.05) & WISE (PI-CCD) \\
55238.52 & 76.52 &   &  &17.42 (0.04) &16.83 (0.03) &16.53 (0.03) & LT (RATCam) \\
55245.81 & 83.81 & 20.49 (0.12) &18.81 (0.15) &17.52 (0.12) &16.90 (0.02) &16.53 (0.04) & NTT (EFOSC) \\
55246.47 & 84.47 &   &18.86 (0.28) &17.45 (0.12) &16.90 (0.04) &16.60 (0.04) & WISE (INDEF) \\
55262.73 & 100.73 & 20.80 (0.25) &18.94 (0.04) &17.71 (0.04) &17.10 (0.03) &16.78 (0.04) & NTT (EFOSC) \\
55268.36 & 106.36 &   &  &17.89 (0.05) &17.23 (0.04) &16.96 (0.04) & LT (RATCam) \\
55270.87 & 108.87 &   &19.44 (0.58) &17.96 (0.17) &17.41 (0.20) &17.08 (0.15) & Mt.Ekar 1.82m (AFOSC) \\
55274.48 & 112.48 &   &19.58 (0.68) &18.07 (0.26) &17.43 (0.11) &17.14 (0.05) & WISE (PI-CCD) \\
55276.59 & 114.59 & 21.74 (0.54) &19.97 (0.15) &18.43 (0.16) &17.75 (0.06) &17.36 (0.05) & NOT (ALFOSC) \\
55277.44 & 115.44 &   &20.13 (0.19) &18.51 (0.05) &17.80 (0.03) &17.46 (0.04) & LT (RATCam) \\
55278.47 & 116.47 &   &  &18.73 (0.05) &17.96 (0.02) &17.60 (0.02) & LT (RATCam) \\
55279.51 & 117.51 &   &20.58 (0.23) &18.96 (0.11) &18.16 (0.04) &17.73 (0.04) & LT (RATCam) \\
55280.54 & 118.54 &   &20.98 (0.39) &19.26 (0.14) &18.40 (0.06) &17.97 (0.04) & LT (RATCam) \\
55282.38 & 120.38 &   &21.45 (0.32) &20.07 (0.06) &18.98 (0.03) &18.51 (0.04) & LT (RATCam) \\
55285.61 & 123.61 &   &  &20.51 (0.15) &19.14 (0.04) &18.71 (0.04) & LT (RATCam) \\
55288.41 & 126.41 &   &22.54 (0.56) &20.53 (0.11) &19.35 (0.03) &18.70 (0.05) & NOT (ALFOSC) \\
55296.52 & 134.52 &   &22.31 (0.28) &20.61 (0.12) &19.50 (0.08) &  & NOT (ALFOSC) \\
55314.48 & 152.48 &   &22.64 (0.21) &20.81 (0.04) &19.68 (0.03) &19.02 (0.03) & NOT (ALFOSC) \\
55343.42 & 181.42 &   &  &21.03 (0.06) &19.90 (0.03) &19.28 (0.04) & NOT (ALFOSC) \\
55349.43 & 187.43 &   &22.82 (0.35) &  &  &  & NOT (ALFOSC) \\ 
\hline
\end{tabular}	 				
\end{center}
\label{t_phot}
\end{table*}

\begin{table*}
\caption{Log of infrared photometric measurements of SN 2009md. Errors are given in paranthesis.}
\begin{center}
\begin{tabular}{lccccr}
\hline
JD-2400000 &Phase (d) &$J$ &$H$ &$K$	&Instrument	\\
\hline
55209.58 & 47.58 & 16.11 (0.10) &15.85 (0.05) &  & LT (SupIRCam) \\
55220.73 & 58.73 & 16.13 (0.04) &15.82 (0.06) &15.70 (0.06) & NTT (SOFI) \\
55246.87 & 84.87 & 16.25 (0.03) &15.93 (0.07) &15.85 (0.07) & NTT (SOFI) \\
55263.60 & 101.60 & 16.51 (0.07) &16.14 (0.07) &  & NTT (SOFI) \\
55305.68 & 143.68 & 18.20 (0.06) &17.55 (0.11) &17.51 (0.11) & NTT (SOFI) \\
55348.41 & 186.41 & 18.53 (0.15) &17.89 (0.16) &17.92 (0.13) & TNG (NICS) \\
55352.88	& 190.88	& $>17.3$		& $>16.0$		&		& LT (SupIRCam) \\
\hline
\end{tabular}	 				
\end{center}
\label{t_nirphot}
\end{table*}

\subsection {Bolometric lightcurve and ejected $^{56}$ Ni mass}
\label{s3b}

\begin{figure}
\includegraphics[width=0.48\textwidth,angle=0]{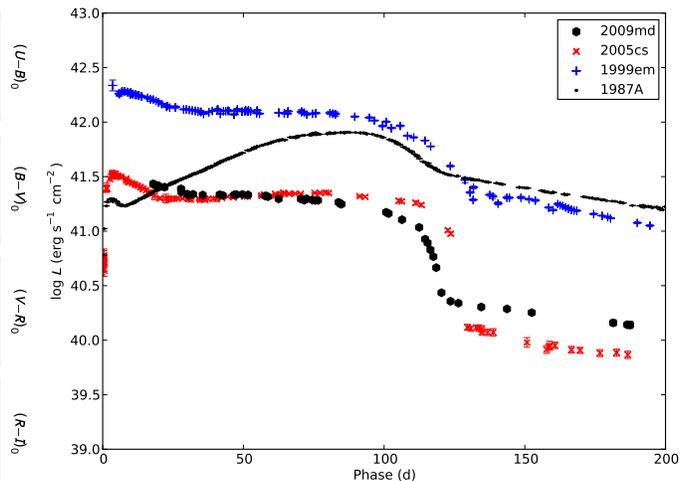}
\caption{Pseudo-bolometric lightcurve for SN 2009md, integrated over the $U$ to $K$ wavelength range. For comparison, we show the sub-luminous Type IIP SN 2005cs, the normal Type IIP SN 1999em, and the peculiar Type IIP SN 1987A.}
\label{f9}
\end{figure}

The pseudo-bolometric $UBVRIJHK$ lightcurves of SNe 2009md, 2005cs, 1999em and 1987A are shown in Fig. \ref{f9}. The difference between these and a true bolometric lightcurve is caused by flux falling outside the covered wavelength range. At early times a substantial fraction of the flux is expected to fall in the ultraviolet (UV). For SN 2005cs, Pastorello et al. (\citealp{Pas09}) showed that 60 per cent of the flux falls in the UV during the first $\sim$20 d, while it is negligible at late times. Assuming the Rayleigh-Jeans law the IR flux redwards of the $K$-band is $\sim$5 per cent during the plateau phase and $\sim$10 per cent during the tail phase. Although we cannot exclude an IR excess we take the last value as an estimate of the error in the pseudo-bolometric lightcurve during the tail phase.

To calculate the pseudo-bolometric lightcurves from the observed absolute magnitudes we interpolate the missing epochs in the filter lightcurves, and for each epoch integrate the total flux over the full wavelength range from $U$ to $K$. To cover missing epochs in the filter lightcurves, we first linearly interpolated the magnitude for the filter with the most data. Missing data in successive neighboring filters were then interpolated by adding/subtracting the linearly interpolated colour at that epoch. This scheme was used as the colour evolution of SNe is usually slower than the evolution in magnitude. NIR magnitudes are badly constrained at early times and during the drop off the plateau. The $U$-band magnitudes are also poorly constrained at very early and late times.

The total flux in the $U$ to $K$ wavelength range could then be calculated. We first log-linearly interpolated the flux per wavelength between the effective wavelengths of the filters. This was done under the constraint that the weighted average over the response functions (Moro \& Munari \citealp{Mor00}) equaled the flux per wavelength as determined by the zero points (Bessell \citealp{Bes79}; Cohen et al. \citealp{Coh03}). The total flux was then integrated between the short and long wavelength half-maximum points of the $U$ and $K$ filters.

Comparing SN 2009md to SN 2005cs (Fig. \ref{f9}), we see that they are indeed very similar although the latter shows a sharper and deeper drop off the plateau. Comparing SNe 2009md and 2005cs to the normal Type IIP SN 1999em, we find that the plateau length is similar, but the plateau and tail luminosity are much lower. It is clear that SN 2009md, as far as the bolometric lightcurve is concerned, is a sub-luminous Type IIP SN.

We estimate the $^{56}$Ni mass by comparing the pseudo-bolometric luminosity of SNe 2009md and 1987A (from the extensive CTIO and SAAO datasets). This assumes that the late-time ejecta are optically thick to the $\gamma$-rays produced by the $^{56}$Co decay, and hence we are seeing all the luminosity from radioactive decay. This can be observationally tested by measuring the decline rate on the tail. The theoretically expected value for fully trapped $\gamma$-rays is 0.0098 mag d$^{-1}.$ We measure the decline rate between 144 and 186 d to be 0.0082 $\pm$ 0.0017 mag d$^{-1}$, which is close but slightly lower than the theoretically expected value. A similar slow decline was also estimated for SN 2005cs. Pastorello et al. (\citealp{Pas09}) tentatively attributed it to the ``tail plateau phase'' described by Utrobin (\citealp{Utr07b}) caused by thermal radiation from the still hot inner ejecta. Comparing the pseudo-bolometric luminosities of SNe 2009md and 1987A at 186 d, and using a ${}^{56}$Ni mass of 0.069 \msun\ (Bouchet et al. \citealp{Bou91}) for SN 1987A, we obtain an ejected $^{56}$Ni mass for SN 2009md of $(5.4\pm1.3) \times 10^{-3}$ \msun.

\subsection{Spectroscopic follow-up}
\label{s3c}
 
Spectroscopic observations of SN 2009md were obtained with the NTT + EFOSC2 and SOFI, the Mt. Ekar 1.8m telescope + AFOSC and the NOT + ALFOSC. For all optical spectroscopic data, the {\sc quba} pipeline was used to reduce the data in a standard manner. Most optical spectra were flux calibrated using spectroscopic standards obtained on the same night as the SN observations, but in a few cases such data were not available and archive sensitivity functions have been used instead. The NTT + SOFI NIR spectra were reduced with the {\sc sofi} pipeline, extracted with the {\sc twodspec} {\sc iraf} package and corrected for telluric absorption using a solar analogue spectroscopic standard and the known NIR spectrum of the Sun. Details of all spectroscopic observations, the telescope and instrument used, epoch and instrument characteristics are given in Table \ref{t_speclog}.

\begin{table*}
\caption{Log of spectroscopic observations of SN 2009md}
\begin{center}
\begin{tabular}{lcccc}
\hline
JD-2400000 & Phase (d) & Range (\AA) & Resolution (\AA) & Instrument\\
\hline
55173.78 & 12 & 3200-9100 & 21 & NOT (ALFOSC) \\
55176.66 & 15 & 3200-9100 & 21 & NOT (ALFOSC) \\
55189.27 & 27 & 3380-10320 & 16 & NTT (EFOSC2) \\
55209.66 & 48 & 3200-9100 & 16 & NOT (ALFOSC) \\
55216.08 & 54 & 3550-10000 & 25	& Mt. Ekar 1.82m (AFOSC)\\
55220.27 & 58 & 9500-16400 & 22 & NTT (SOFI) \\
55245.28 & 83 & 3380-10320 & 16 & NTT (EFOSC2) \\
55262.17 & 100 & 3380-10320 & 16 & NTT (EFOSC2) \\
55263.16 & 101 & 9500-16400 & 22 & NTT (SOFI) \\
55263.21 & 101 & 15300-25200 & 34 & NTT (SOFI) \\
55271.97 & 110 & 3550-10000 & 25	& Mt. Ekar 1.82m (AFOSC)\\
55304.17 & 142 & 3685-9315 & 21 & NTT (EFOSC2) \\
\hline
\end{tabular}
\end{center}
\label{t_speclog}
\end{table*}

\begin{figure}
\includegraphics[width=0.5\textwidth,angle=0]{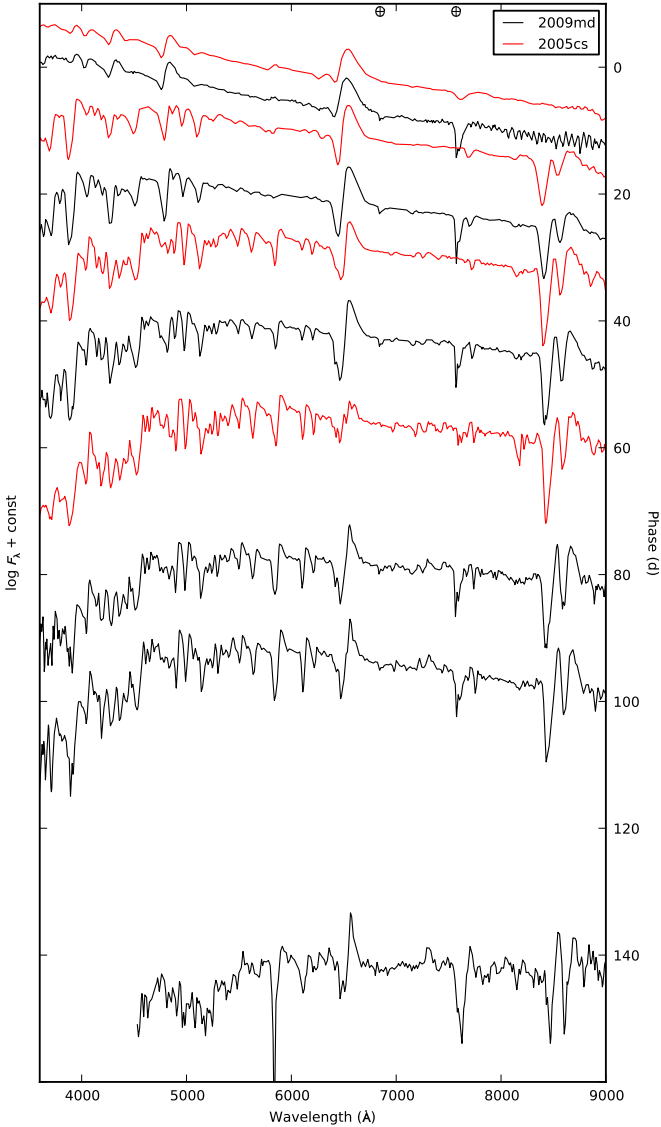}
\caption{Time series of spectra for SNe 2009md (black) and 2005cs (red). To visualise the temporal evolution the spectra have been aligned to the time axis at the right border of the panel. The spectra have been corrected for redshift and dereddened. Telluric features are indicated with a $\oplus$ symbol. The lower signal-to-noise spectra of SN 2009md from the Mt. Ekar 1.82m telescope and at 15 d from NOT have been excluded for clarity.}
\label{f11}
\end{figure}

\begin{figure}
\includegraphics[width=0.5\textwidth,angle=0]{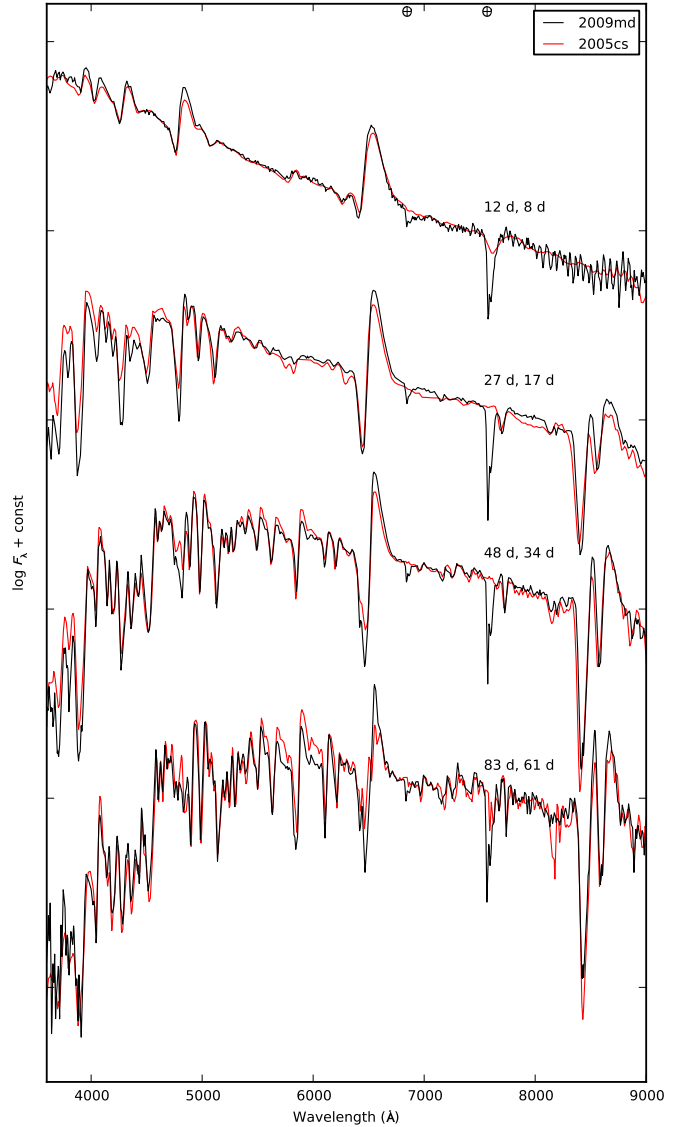}
\caption{Overplots of the epochs shown in Fig. \ref{f11} giving the best match in SNID (Blondin \& Tonry \citealp{Blo07}). The spectra have been corrected for redshift and dereddened. Telluric features are indicated with a $\oplus$ symbol.}
\label{f12}
\end{figure}

The optical spectral evolution from $12-142$ d of SN 2009md is shown in Fig. \ref{f11}, together with that of SN 2005cs (overplotted in red). It is clear from this figure and the detailed comparisions in Fig. \ref{f12} that SN 2009md is also spectroscopically very similar to SN 2005cs. The spectral evolution very much follows that of SN 2005cs as described in Pastorello et al. (\citealp{Pas09}). At early epochs, the spectrum has a blue continuum, with the dominant features being H and He lines together with some evidence of Si {\sc ii} and Fe {\sc ii}. Around $10-30$ d post explosion, the O {\sc i} $\lambda7774$, Ca {\sc ii} H \& K lines and NIR triplet, Na {\sc i} D and numerous strong Fe {\sc ii} lines appear. At later times the continuum becomes redder and a number of other metal lines appear, such as Ti {\sc ii}, Sc {\sc ii} and Ba {\sc ii}, and absorption features grow deeper. As the line widths in sub-luminous Type IIP SNe are smaller than in normal Type IIP SNe, lines that are normally blended may be resolved. This may explain the Ba {\sc ii} line which appears in the blue wing of the H$\alpha$ line and other features not seen in normal Type IIP SNe spectra.

\begin{figure}
\includegraphics[width=0.5\textwidth,angle=0]{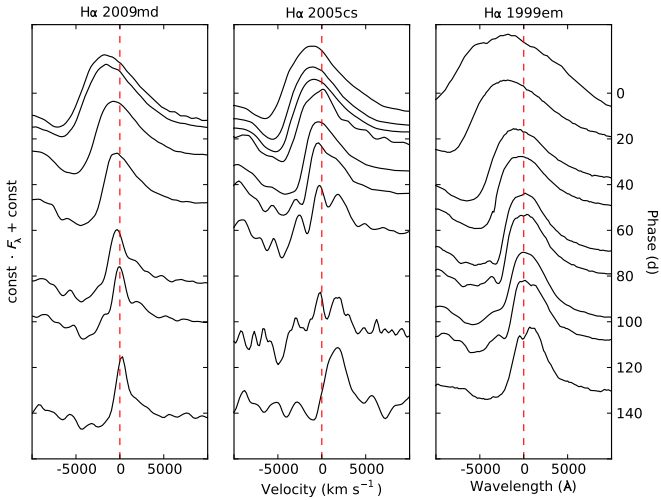}
\caption{Time evolution of the H$\alpha$ line in SN 2009md (left panel), as compared to SNe 2005cs (center panel) and 1999em (right panel). To visulize the temporal evolution the spectra have been aligned to the time axis at the right border of each panel. The spectra have been corrected for redshift and dereddened. The emission peak is noticeably blue-shifted at earlier epochs.}
\label{f13}
\end{figure}

\begin{figure}
\includegraphics[width=0.5\textwidth,angle=0]{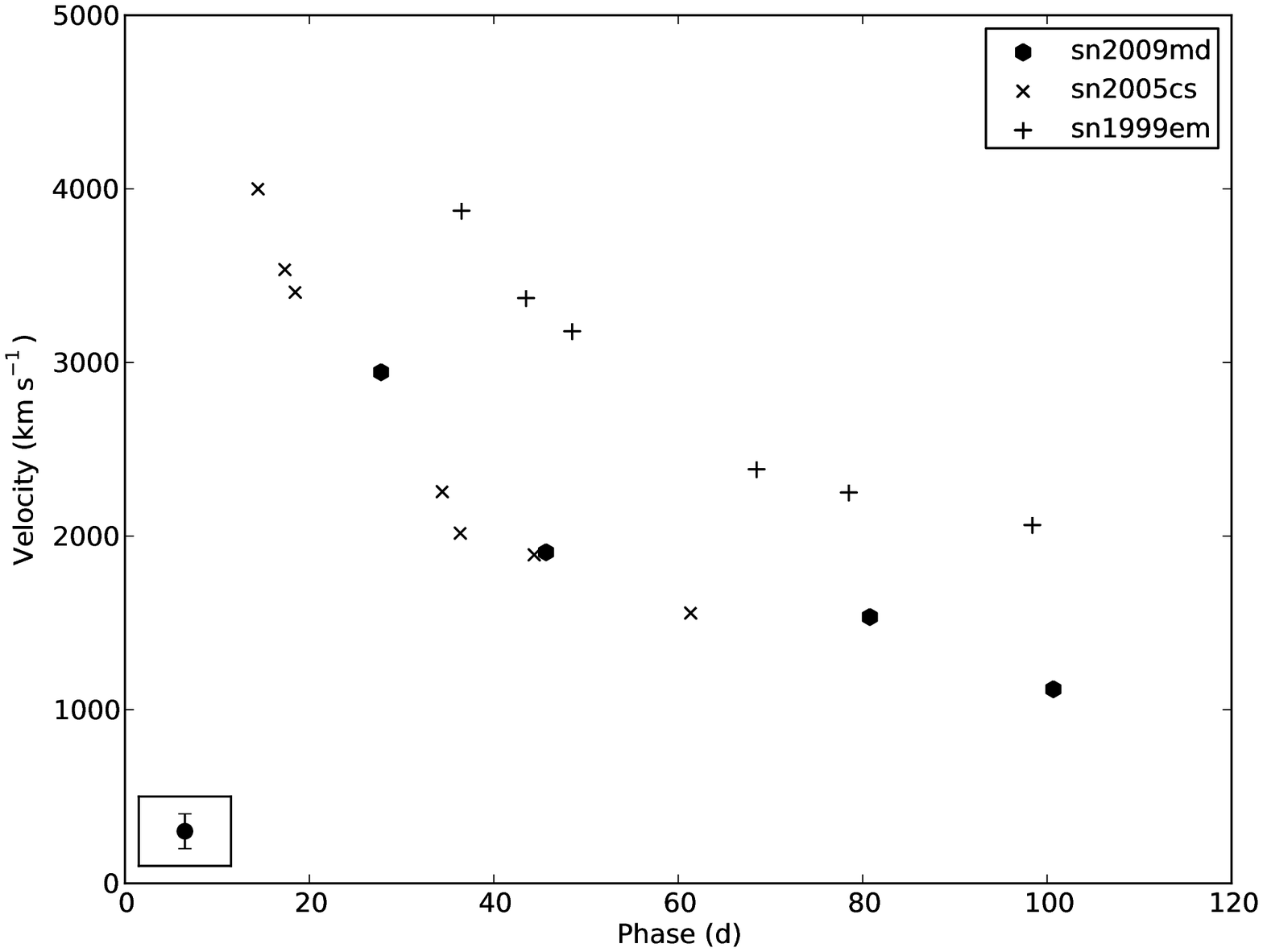}
\caption{Photospheric velocity evolution of SN 2009md as measured from the Sc {\sc ii} 6246 \AA\ line, compared to those of SNe 2005cs and 1999em. The typical measurement error is indicated in the lower left corner.}
\label{f14}
\end{figure}

Despite the similarity to SN 2005cs there are also some noticable differences. In Fig. \ref{f13} we show the evolution of the H$\alpha$ line as compared to SNe 2005cs and 1999em. It can be seen that the H$\alpha$ evolution between 60 and 120 d is rather different, with the line appearing weaker at later epochs in SN 2005cs as compared to SN 2009md. Comparing the time series of spectra of SN 2009md with SN 2005cs in Fig. \ref{f11}, the former appears to be evolving more slowly than the latter; this is also confirmed by cross-correlation of the spectra using SNID (Blondin \& Tonry \citealp{Blo07}). At $\sim$10 days the match is quite good, at 27 d the best match is 17 d, at 48 d the best match is 34 d, and at 83 d the best match is 61 d, where the former epochs refer to SN 2009md and the latter to SN 2005cs.

In Fig. \ref{f14} the velocity evolution of the Sc {\sc ii} $\lambda$6246 line as measured from the P-Cygni minimum is shown. This line is believed to be a good indicator of the photospheric velocity (Maguire et al. \citealp{Mag10b}). As seen in Fig. \ref{f14} the line velocities are considerably lower for SNe 2009md and 2005cs than for the normal Type IIP SN 1999em. We have already shown that SN 2009md satisfies two of the criteria of the sub-luminous Type IIP SNe, namely a faint absolute magnitude, and a low ejected $^{56}$Ni mass. As low line velocities is the third signature we can thus classify SN 2009md as a sub-luminous Type IIP SN.

\begin{figure}
\includegraphics[width=0.35\textwidth,angle=270]{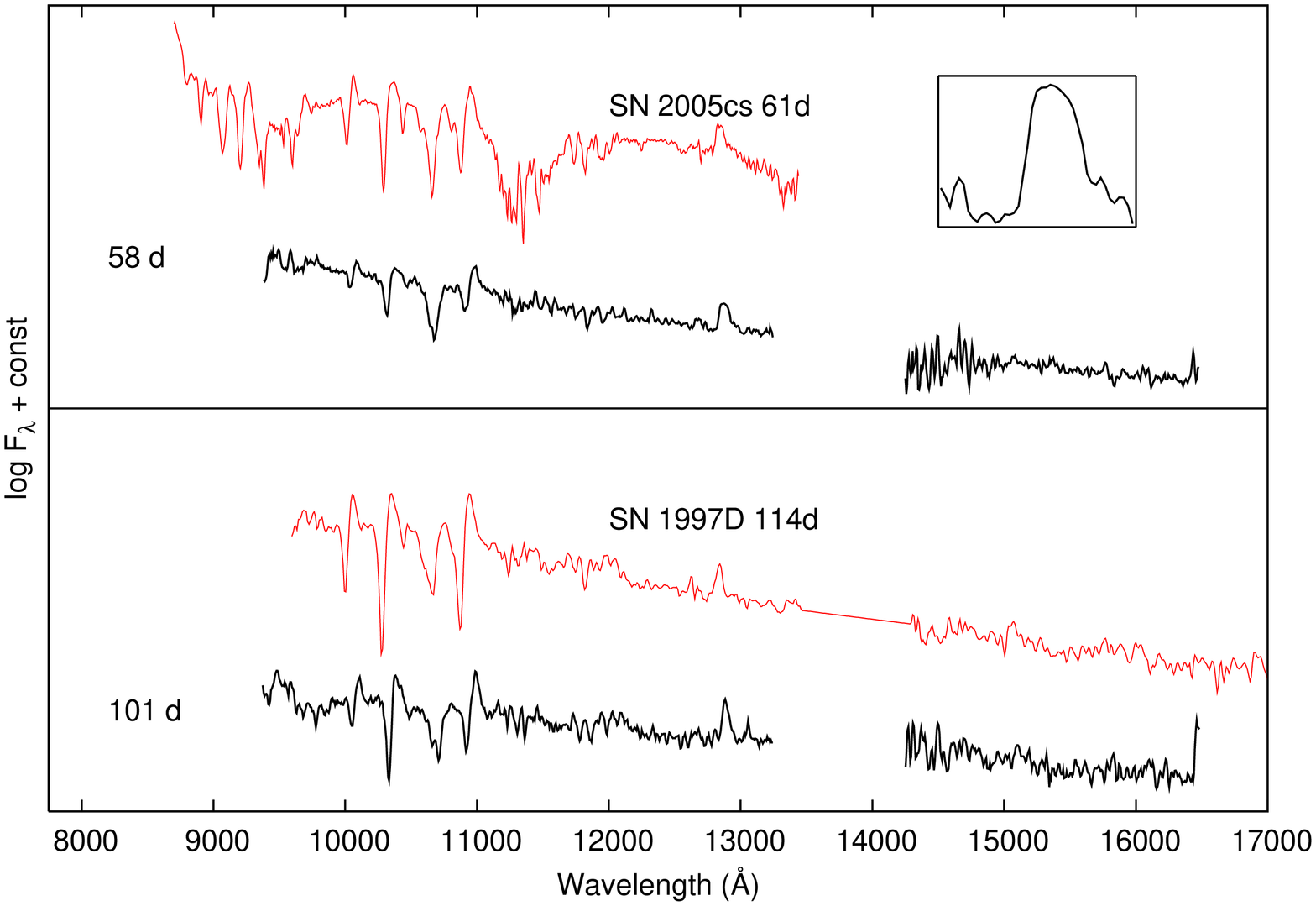}
\caption{De-reddened and redshift corrected near infrared spectra of SN 2009md. The upper panel shows SN 2009md (in black) at the midpoint of the plateau phase, as compared to SN 2005cs at a similar epoch. The second panel shows SN 2009md at the end of the plateau, as compared to the sub-luminous SN 1997D. The box in the upper panel shows a magnified view of the shape of the Pa$\beta$ line at 58 d as discussed in the text.}
\label{f16}
\end{figure}

Two epochs of NIR spectra of SN 2009md were obtained with NTT + SOFI, the first at 58 d and the second at 101 d (Fig. \ref{f16}). Ideally we would compare these to the same SN as in the optical at a similar epoch, unfortunately the dearth of NIR spectra precludes this. Hence we have compared the 58 d spectrum to SN 2005cs and the 101 d spectrum to SN 1997D.

At 58 d, the NIR spectrum of SN 2009md displays the series of Paschen lines from H. Pa$\beta$ at 1.28 $\mathrm{\upmu m}$ appears only in emission, and appears to have a somewhat boxy profile. Boxy line profiles are usually associated with an optically thin line (eg. Fransson \citealp{Fra84}). It is hence possible that the Pa$\beta$ emission we observe is formed further out in the ejecta. Pa$\gamma$ and Pa$\delta$ at 1.09 and 1.00 $\mathrm{\upmu m}$ respectively have P-Cygni profiles. The Pa$\gamma$ line is blended with He {\sc i} at 1.083 $\mathrm{\upmu m}$.

\section{Discussion}
\label{s4}

In the following, we use the standard candle method for Type IIP SNe to verify the distance to NGC 3389. We attempt to explain the discrepancy between the predictions of stellar evolutionary models and the observed luminosity of sub-luminous Type IIP SN progenitors. The mass range of progenitors which produce sub-luminous Type IIP SNe is estimated both from progenitor detections and SN rates. We also discuss the relationship of the ejected $^{56}$Ni mass to the progenitor luminosity.

\subsection{Verifying the distance to SN 2009md}
\label{s4a}

To verify the distance to the host galaxy of SN 2009md, NGC 3389, we use the standard candle method for Type IIP SNe (Hamuy \& Pinto \citealp{Ham02}; Poznanski et al. \citealp{Poz09}; D'Andrea et al. \citealp{Dan10}). Hamuy \& Pinto (\citealp{Ham02}) show that the $V$-band magnitude and expansion velocity of Type IIP SNe satisfy the equation

\begin{equation}
V_\mathrm{p} - A_V + 6.504(\pm0.995) \mathrm{log}({{v_\mathrm{p}}\over{5000}}) = 5 \mathrm{log} (cz) - 1.294(\pm0.131)
\end{equation}

where $V_\mathrm{p}$ is the apparent $V$-band magnitude at 50 d post explosion, $v_\mathrm{p}$ is the expansion velocity in \kms\ at the same epoch, $A_V$ is the $V$-band extinction in magnitudes, and $z$ is the redshift of the host galaxy. Taking the magnitude of SN 2009md at 50 d from Table \ref{t_phot}, $V_\mathrm{p}$ = 17.30$\pm$0.03 mag, and the expansion velocity of $\sim$2000 \kms\ from the spectrum of SN 2009md closest to 50 d (Fig. \ref{f14}), we find a host galaxy redshift $z = 0.0046$, corresponding to a distance modulus of 31.38 mag. This is in good agreement with the distance estimates presented in Section \ref{s1}.

Maguire et al. (\citealp{Mag10b}) have extended the technique to the NIR, where they find a trend of reduced dispersion compared to optical wavelengths in the resultant Hubble Diagram. Modifying eq. 1 of Maguire et al. to solve for the distance modulus

\begin{equation}
\mu = m_J - J_{0} + \alpha \mathrm{log}_{10}  {{v_\mathrm{{Fe\,{II}}}}\over{5000}}-R_{J}[A_V-A_J]
\end{equation}

and using the Sc\,{\sc ii} velocity of 2000 \kms\ and $m_J = 16.112 \pm 0.097$ mag from the spectrum and NIR photometry at $\sim$ 50 d, together with the values of $J_{0}=-18.06\pm0.25$ mag, $R_J = 0.108$ and $\alpha = 6.33\pm1.2$ (where $\alpha$ is a fitting constant) from Maguire et al., we obtain a distance modulus $\mu = 31.63\pm0.55$ mag, which again is in agreement with the distance presented in Section \ref{s1}, $\mu = 31.64\pm0.21$.

\subsection{Producing faint progenitors with evolutionary models}
\label{s4b}

The classical picture of stellar evolution sees stars with a main sequence mass $\lesssim$ 8 \msun\ form white dwarfs, stars with $8 \lesssim M \lesssim 11$ \msun\ become Super-Asymptotic Giant Branch (SAGB) stars which may lead to electron-capture SNe (ECSNe), and stars with $M \gtrsim 11$ \msun\ burning through to iron core-collapse (see, for example, Woosley, Heger \& Weaver \citealp{Woo02}). Unfortunately, the mass range over which these changes occur is small (and indeed comparable to the uncertainty in our progenitor mass). Furthermore, the precise mass thresholds for each of the three regimes are sensitive to the details of the evolutionary models used; with the most uncertain aspects of stellar evolution, namely convection and mixing, nuclear reaction rates, and the neutrino cooling rate, being most significant for this regime. In recent years, however, there have been several detections of low luminosity progenitors, which lead us to question the accuracy of this picture (Smartt et al. \citealp{Sma09b}).

\begin{figure}
\includegraphics[width=0.35\textwidth,angle=270]{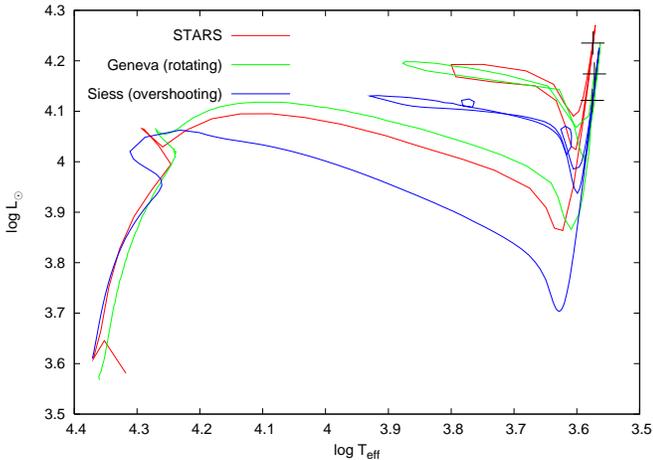}
\caption{Comparison of evolutionary codes at the end of core He burning (the last stage all three codes are evolved through). All models are for a star with an initial mass of 9 \msun, at solar metallicity. The point of core He exhaustion is marked for all three tracks with a + symbol.}
\label{f17}
\end{figure}

There is a lack of stellar evolutionary models which follow stars at the low end of the mass range of Type IIP progenitors through, and beyond, core helium burning. This is significant as there are differences between the various codes, not only in terms of the numerical techniques used, but also their treatment of some of the more uncertain physics in the models, eg. mixing related phenomena such as rotation and convective overshooting. To examine the differences between some of the available codes, we took a 9 \msun\ model at solar metallicity (Z=0.020), and examined its luminosity at the end of core He burning (which we have defined as when the core He fraction by mass drops below 1 per cent). We have plotted the luminosity against temperature for the rotating Geneva (Meynet \& Maeder \citealp{Mey03}), STARS (Eldridge \& Tout \citealp{Eld04}) and Siess (\citealp{Sie06}) codes in Figure \ref{f17}. All three codes are in agreement to within 0.1 dex in the luminosity at the end of core He burning. On the basis of this, we can conclude that while there exists a need for models that follow stellar evolution to later stages, the problems discussed in the following are not model-specific, and can instead be attributed to a more general problem with our understanding of stars in this mass range.

AGB and SAGB stars arise after several dredge-ups in between the major nuclear burning stages of stellar evolution. After the star forms a He core, it will ascend the red giant branch. This is accompanied by a deepening of the stars convective envelope that mixes nuclear processed material to the surface of the star during the first dredge-up. This process terminates  with the ignition of core He burning. The process occurs again at the end of core He burning, in a second dredge-up. As before, a deep convective envelope penetrates into the star, and brings the majority of the He which surrounds the carbon-oxygen (CO) core to the surface. The star is then left with a CO core, essentially a CO white dwarf, surrounded by thin He- and H-burning shells. Such an arrangement is unstable, and so burning progresses in a series of thermal pulses (eg. Stancliffe, Tout \& Puls \citealp{Sta04}). In the most massive cases the CO core is massive enough that carbon is ignited and the CO core is processed into an oxygen-neon (ONe) core. Such a star is called a Super-AGB star. If the ONe core reaches the Chandrasakhar mass then an electron capture core-collapse will occur (Eldridge \& Tout \citealp{Eld04}; Siess \citealp{Sie07}; Poelerands \citealp{Poe08}; Pumo et al. \citealp{Pum09}).

The fact that the progenitor identified for SN 2009md has a luminosity of $\sim$4.54 dex indicates that this star has not undergone the second dredge-up, and hence is not an SAGB star. Furthermore, there have now been several detections of progenitors with similarly low luminosities (eg. SNe 2005cs, 2003gd \& 2008bk). The problem of the low luminosities has been discussed by Smartt et al. (\citealp{Sma09b}), for a full discussion of the issues raised the reader is referred to this work.

\begin{figure*}
\includegraphics[width=1\textwidth,angle=0]{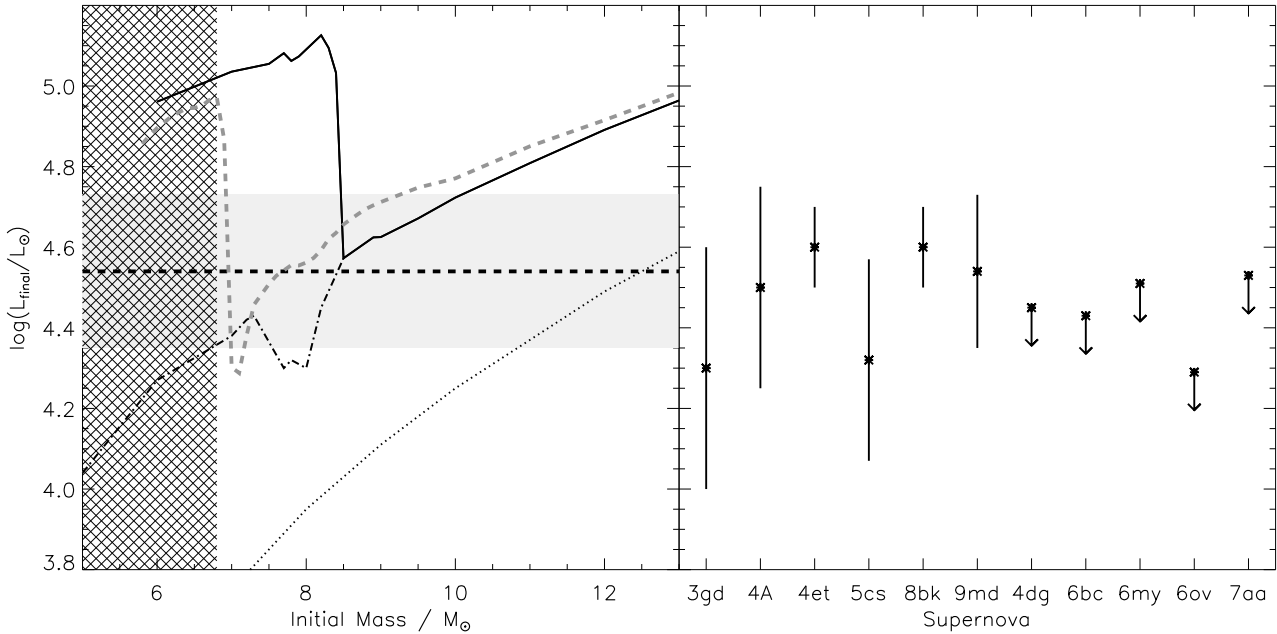}
\caption{The luminosity of models as a function of mass, similar to Fig. \ref{f5}, but showing the effects of enhanced carbon burning. In the left panel, the solid black line is the final luminosity of the standard STARS model, the dashed grey line is the final luminosity of the STARS model with an enhanced $^{12}$C burning rate. The dash-dot line is the luminosity of the standard STARS models immediately prior to second dredge-up, and the dotted line is the standard model luminosity at end of core helium burning (which is the same for both normal and enhanced carbon burning models). We indicate the region in which stars will not explode as core-collapse SNe with a hatched area. In the right panel, we show progenitor luminosities from detections and upper limits.}
\label{f18}
\end{figure*}

To explain the fact that we now have several SNe where the progenitor is less luminous than expected from evolutionary models, we are left with two possibilities; either prevent the second dredge-up from occurring, or ensure that the star evolves rapidly enough that it explodes before the second dredge-up. The former option is more difficult. Although convection related phenomena are some of the least well understood aspects of stellar evolution models, it is difficult to find a physical process that will prevent convection in and around the core and thus prevent second dredge up. We thus turn to attempting to force stellar evolution to proceed faster. The one key detail that may be most uncertain in current models is the carbon-burning rate. Recent measurements of the $^{12}$C cross-section suggest that a resonance may be present that boosts the burning rate by a factor of $10^5$ (Spillane et al. \citealp{Spi08}). To investigate the effect of this new higher rate, we have created a series of stellar models similar to those discussed in Eldridge \& Tout (\citealp{Eld04}) with the carbon-burning rate boosted by such a factor. We find that the increase in the $^{12}$C burning rate decreases the minimum mass for a SN by one solar mass to 7 \msun\, but except for a narrow mass range, the final luminosity of the models does not decrease, and so the problem remains (see Figure \ref{f18}).

It is difficult to follow these models with enhanced carbon burning though to the formation of an iron core, as the formation of  a very hot, thin shell around the degenerate core causes numerical difficulties. However, the high temperatures reached in the core towards the end of the models (10$^{9.3}$ K) suggests that these models may in fact form an Fe core. This is helped by the avoidance of the second dredge-up, as the core does not become dense enough for significant cooling through neutrino emission before the latter burning stages. However, we must once again stress that this is an exploratory \emph{attempt} to produce a model which avoids second dredge-up, and may, or may not, be correct.

\subsection{The mass range of the progenitors of sub-luminous Type IIP SNe}
\label{s4c}

If we assume two channels for sub-luminous Type IIP SNe, a ``low mass''  channel between an arbitrary mass range of 8--10 \msun\, and a ``high mass'' channel between an arbitrary mass range of 23--27 \msun, then for a standard Salpeter initial mass function (IMF) with a slope of $-2.35$, $\sim$5 times as many sub-luminous SNe will come from the low mass channel as from the high mass channel. In this context, the fact that the three detected sub-luminous SN progenitors (for SNe 2008bk, 2005cs and 2009md) are all low mass is not surprising.

\begin{figure}
\includegraphics[width=0.48\textwidth,angle=0]{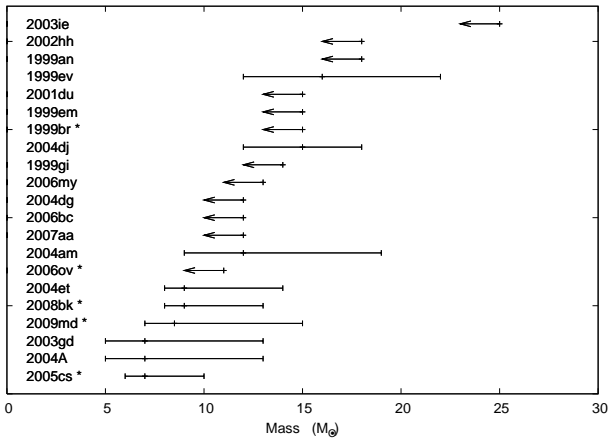}
\caption{Measured masses for the sample of SNe from Smartt et al. (\citealp{Sma09b}) with progenitor detections or upper mass limits. SN 2009md is also added to this figure. Sub-luminous Type IIP SNe are marked with an asterisk, and have had their masses or mass limits recalculated using synthetic photometry of MARCS models as discussed previously.}
\label{f20}
\end{figure}

While we can clearly say from the progenitors of sub-luminous Type IIP SNe found so far that at least a significant fraction of them come from low mass progenitors, as shown in Fig. \ref{f20}, it is of interest to try and restate this in more quantitative terms. Hence we have followed the methodology of Smartt et al. (\citealp{Sma09b}) in calculating the probable maximum and minimum mass for Type IIP core collapse SNe (for a detailed description of the methodology the reader is referred to this paper). In brief, we have two free parameters, $M_{\mathrm{min}}$ and $M_{\mathrm{max}}$ which are  the minimum and maximum masses for SN progenitors respectively. We calculate the likelihood of a progenitor having a mass with respect to the IMF, maximizing this over all progenitors and upper limits. An important point to note is that only \emph{detections} are taken into account when calculating the lower mass limit, this is because the non-detections of progenitors leave the lower mass unconstrained, and due to the nature of the IMF, these will tend towards lower masses. We are also assuming that sub-luminous Type IIP SNe come exclusively from a single continuous mass range.

We have three progenitor detections, and two upper limits on mass. To ensure a consistent approach, we have recalculated the progenitor masses of SNe 2005cs, 2008bk, 2005ov and 1999br using the synthetic photometry approach discussed in Section \ref{s2d}. We find that the upper limit for the progenitor of SN 1999br is unchanged at $<15$ \msun, SN 2006ov increases by to $<11$ \msun, and the progenitor masses of SNe 2005cs and 2008bk remain the same  (although the uncertainty in the mass of the former is increased). We calculate the most probable progenitor mass range for the sub-luminous Type IIP SNe to be between 7.5 and 9.5 \msun, although as this is based on so few data points the significance of this result is low.

As a consistency check, taking all SNe in the volume and time limited sample of Smartt et al. (\citealp{Sma09b}) (within 20 Mpc, discovered between Jan 1998 and June 2008) we can calculate the relative rate of sub-luminous SNe as a fraction of all Type IIP SNe to be 23 per cent. Using this, and assuming upper and lower mass limits for Type IIP SN progenitors of 8.5 and 16.5 \msun\ respectively (Smartt et al. \citealp{Sma09b}) and a Salpeter IMF, we find the maximum mass for a sub-luminous Type IIP progenitor to be 9.3 \msun\, which is in accord with that the mass range suggested previously.

\subsection{Explosion energy and ejecta mass}
\label{s4d}

Using simple scalings, we can compare the explosion energy $E$ and ejecta mass $M$ of SN 2009md to those of SN 1999em. Eq. 1 and 2 in Kasen \& Woosley (\citealp{Kas09}) give scalings for $E$ and $M$ in terms of the plateau luminosity $L_\mathrm{sn}$, duration $t_\mathrm{sn}$, ejecta velocity $v_\mathrm{sn}$ and progenitor radius $R$,
 
\begin{equation}
E \propto {{L_\mathrm{sn}t_\mathrm{sn}^{2}v_\mathrm{sn}}\over {R}}
\end{equation}
 
\begin{equation}
M \propto {{L_\mathrm{sn}t_\mathrm{sn}^{2}\over {Rv_\mathrm{sn}}}}
\end{equation}

From Fig. 9 we see that the plateau lengths are similar, while the plateau luminosity is a factor of $\sim$7 lower for SN 2009md. From Fig. 13 we see that the ejecta velocity is $\sim$1.5 lower for SN 2009md. The radius of SN 2009md is estimated to be $\sim$500 R$_\odot$ from the effective temperature and luminosity given in Section 2.6, which is comparable to the value found for SN 1999em by Utrobin (\citealp{Utr07b}). This gives a factor of $\sim$10 lower energy and a factor of $\sim$5 lower mass. Although a crude estimate, this clearly suggests a very low explosion energy and a low ejecta mass as compared to normal Type IIP SNe.

\subsection{The production of $^{56}$Ni}
\label{s4e}

The $^{56}$Ni synthesized in SNe, which powers the late-time luminosity after the plateau stage, is largely formed by the explosive burning of oxygen and silicon. As noted by Eldridge, Mattila \& Smartt (\citealp{Eld07}) and Kitaura, Janka \& Hillebrandt (\citealp{Kit06}), an SAGB progenitor will have little oxygen and silicon surrounding the core, and hence we may expect a SAGB star to produce a low mass of ejected $^{56}$Ni. But a low mass ($\sim$9 \msun) progenitor of an Fe core-collapse SN will also produce a small quantity of ejected $^{56}$Ni for the same reason. A high mass ($\sim$25 \msun) progenitor can also lead to the apparent formation of small quantities of $^{56}$Ni, as the $^{56}$Ni that is synthesized falls back onto the nascent black hole (eg. Woosley \& Weaver \citealp{Woo95}; Heger et al. \citealp{Heg03}). 

The low mass of ejected $^{56}$Ni seen in SN 2009md is extremely low when compared to that of a normal Type IIP, albeit twice as high as the $3 \times 10^{-3}$ \msun\ $^{56}$Ni mass estimated for SN 2005cs by Pastorello at al. (\citealp{Pas09}).

\begin{figure}
\includegraphics[width=0.32\textwidth,angle=270]{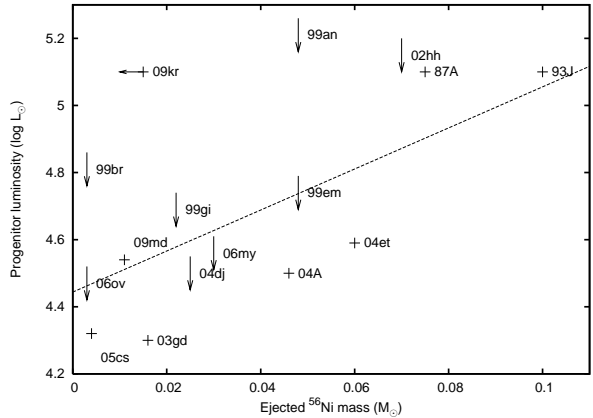}
\caption{Ejected $^{56}$Ni mass against progenitor luminosity (plus symbol) or luminosity limit (arrow), for the sample of Type IIP SNe with both measured progenitor luminosities or limits, and Ni$^{56}$ masses. A least squares fit to the measured masses is also shown with a dashed line.}
\label{f22}
\end{figure}

We have plotted the ejected $^{56}$Ni mass against progenitor luminosity for a sample of SNe in Fig. \ref{f22}. The degree to which $^{56}$Ni mass and progenitor luminosity correlate is striking, although given that the ejected nickel mass is dependent on the final core mass, and the final core mass will be dependent on the initial mass of the star and hence linked to luminosity, perhaps not so surprising. Even the peculiar Type IIP SN 1987A, and SN 1993J, for which a binary scenario was proposed, follow the relation well.

\section{Conclusions}
\label{s5}

To understand SNe and their progenitors, it is necessary to take a holistic approach, and draw together observational data on both the progenitor and its explosion, together with the theoretical framework to interpret these data.

We find a faint, red source in archival images coincident with SN 2009md. We find a best match to a low mass red supergiant, with $M=8.5^{+6.5}_{-1.5}$ \msun, and favor this as the progenitor. While there is some uncertainty as to how the final stages of low mass SN progenitors evolve (as discussed in Section \ref{s4b}), we can be confident that neither a SAGB star, nor a high mass star is the progenitor of SN 2009md. We have made some tentative attempts to evolve a series of low mass models using the STARS code to match our observations, and find that by boosting the carbon burning rate we can alter the final luminosity of models, and lower the minimum mass for the models to not undergo second dredge-up. We stress however, that this is only a first attempt, and that there is a need to both evolve the codes closer to core-collapse, and to establish whether the enhanced C burning we suggest actually occurs.

This is the third direct detection of the progenitor of a sub-luminous Type IIP, confirming that at least a significant fraction of this class of SNe comes from low mass ($\sim 8$ \msun) stars. It is tempting to suggest a scenario in which all these SNe come from low mass progenitors, and are produced by low energy Fe core-collapse explosions which eject small amounts of $^{56}$Ni. However, we can not rule out the presence of a population of high mass fallback SNe, although we can confirm that if such a population does exist, we have yet to observe its progenitors.

The evolution of SN 2009md is remarkably similar to that of the sub-luminous Type IIP SN 2005cs, with the main difference being the apparently faster spectroscopic evolution of the latter. In terms of absolute magnitude and line velocities, however, we can regard SN 2009md as SN 2005cs's twin.

In this paper, we have presented data from the first $\sim$6 months of SN 2009md. In many ways though, the most interesting work remains to be done. We have presented a challenge, to model a supernova progenitor with a low luminosity immediately prior to core collapse. Such a challenge is far from simple, as the ever more rapid burning of successively heavier elements, combined with dynamical processes such as convection and dredge-up, conspire to make the final years of a star's life the hardest to model. We have also presented a challenge to attempt to model the explosion of a low mass star with an Fe core. If our suggestion that the core of low mass progenitors may in fact become hot enough to burn through to Fe (based on our enhanced C burning models) is correct, then it is of great interest to model an Fe core collapse in a low mass star.

Observationally, there is also future work to be done. Some years hence, deep images of the site of SN 2009md with either an 8 m class telescope, the \emph{HST} or (future) \emph{James Webb Space Telescope} can confirm the disappearance of the progenitor (e.g. Maund \& Smartt \citealp{Mau09}; Gal-Yam \& Leonard \citealp{Gal09}). Accurate measurements of any remaining flux at the progenitor coordinates may also reduce the uncertainties in progenitor mass and luminosity. Another interesting, but less direct, test of the progenitor scenario we have proposed for SN 2009md may be to measure the oxygen abundances in late time nebular spectra. Kitaura, Janka \& Hillebrandt (\citealp{Kit06}) propose such a test to distinguish between low and high mass progenitors, as the former should eject considerably less oxygen (perhaps one hundredth to one thousandth of the mass).

\section{Acknowledgements}

MF is funded by the Northern Ireland Department of Employment and Learning. This work was conducted as part of a EURYI scheme award (esf.org/euryi). The Oskar Klein Centre (ME, JS) is funded by the Swedish Natural Science Research Council. SB, FB and MT are partially supported by the PRIN-INAF 2009 with the project ``Supernovae Variety and Nucleosynthesis Yields''. The Dark Cosmology Centre is funded by the Danish National Research Foundation. AG's work is supported by grants from the Israeli Science Foundation, an FP7 Marie Curie IRG fellowship, a research grant from the Peter and Patricia Gruber Awards, the Weizmann-Minerva program, and the Benoziyo Center for Astrophysics.  

Based on observations obtained at the Gemini Observatory (GN-2010A-Q-54, PI: Crockett \& Gal-Yam), which is operated by the Association of Universities for Research in Astronomy, Inc., under a cooperative agreement with the NSF on behalf of the Gemini partnership: the National Science Foundation (United States), the Science and Technology Facilities Council (United Kingdom), the National Research Council (Canada), CONICYT (Chile), the Australian Research Council (Australia), MinistŽrio da Cincia e Tecnologia (Brazil)  and Ministerio de Ciencia, Tecnolog'a e Innovaci—n Productiva  (Argentina). Based on observations made with the NASA/ESA Hubble Space Telescope, obtained from the data archive at the Space Telescope Institute. STScI is operated by the association of Universities for Research in Astronomy, Inc. under the NASA contract NAS 5-26555.

Based on data collected at the Canada-France-Hawaii Telescope. Based on observations made with the Nordic Optical Telescope, operated on the island of La Palma jointly by Denmark, Finland, Iceland, Norway, and Sweden, in the Spanish Observatorio del Roque de los Muchachos of the Instituto de Astrofisica de Canarias. Partially based on observations made with the Italian Telescopio Nazionale Galileo (TNG) operated on the island of La Palma by the Fundaci—n Galileo Galilei of the INAF (Istituto Nazionale di Astrofisica) at the Spanish Observatorio del Roque de los Muchachos of the Instituto de Astrofisica de Canarias. The Liverpool Telescope is operated on the island of La Palma by Liverpool John Moores University in the Spanish Observatorio del Roque de los Muchachos of the Instituto de Astrofisica de Canarias with financial support from the UK Science and Technology Facilities Council. The Faulkes Telescope North is operated by the Las Cumbres Observatory Global Telescope Network (Hawaii). Based on observations collected at the European Organisation for Astronomical Research in the Southern Hemisphere, Chile (Proposal 184.D-1140). LAIWO, a wide-angle camera operating on the 1-m telescope at the Wise Observatory, Israel, was built at the Max Planck Institute for Astronomy (MPIA) in Heidelberg, Germany, with financial support from the MPIA, and grants from the German Israeli Science Foundation for Research and Development, and from the Israel Science Foundation. Partially based on observations collected at Asiago observatory and Calar Alto Observatory. We especially thank the support staff and telescope operators involved in these observations for their skill, assistance and advice.

This work has been greatly facilitated and expedited by the European supernova collaboration involved in ESO-NTT large program 184.D-1140 led by Stefano Benetti.

This research has made use of the NASA/IPAC Extragalactic Database (NED) which is operated by the Jet Propulsion Laboratory, California Institute of Technology, under contract with the National Aeronautics and Space Administration. We acknowledge the usage of the HyperLeda database (http://leda.univ\-lyon1.fr). This research used the facilities of the Canadian Astronomy Data Centre operated by the National Research Council of Canada with the support of the Canadian Space Agency. Some of the data presented in this paper were obtained from the Multimission Archive at the Space Telescope Science Institute (MAST). STScI is operated by the Association of Universities for Research in Astronomy, Inc., under NASA contract NAS5-26555. Support for MAST for non-HST data is provided by the NASA Office of Space Science via grant NNX09AF08G and by other grants and contracts.

We thank P.-A. Duc for kindly providing his proprietary CFHT+Megacam images of NGC 3389, and Giuliano Pignata for information on SN 2008bk. We also thank Luca Zampieri and Eddie Baron for useful discussions and advice. We thank the anonymous referee for the comments on the paper.

\bsp

\label{lastpage}

\end{document}